\documentclass[12pt]{article}
\usepackage{geometry}
\usepackage{graphicx} 
\usepackage{float}
\usepackage{nameref}  
\usepackage{amsmath}
\usepackage{authblk}
\usepackage{comment}
\usepackage{xcolor}
\usepackage{url}

\usepackage{rotating}  
\usepackage{booktabs}  
\usepackage{multirow}  
\newcommand{\clean}{Clean}
\newcommand{\Bonly}{B Only}
\newcommand{\Oonly}{O Only}
\newcommand{\BO}{B + O}

\title{Atomistic Mechanisms of Oxidation and Chlorine Corrosion in Ni-Based Superalloys: The Role of Boron and Light Interstitial Segregation}

\author[1,2]{Tyler D. Dole\v{z}al\footnote{corresponding authors: tyler.dolezal.1@us.af.mil; rodrigof@mit.edu; liju@mit.edu}}
\author[1]{Rodrigo Freitas$^*$}
\author[1,3]{Ju Li$^*$}
\affil[1]{Department of Materials Science and Engineering, Massachusetts Institute of Technology, Cambridge, MA, USA}
\affil[2]{Department of Engineering Physics, Air Force Institute of Technology, Wright-Patterson Air Force Base, OH, USA}
\affil[3]{Department of Nuclear Science and Engineering, Massachusetts Institute of Technology, Cambridge, MA, USA}

\begin{document}

\maketitle

\begin{abstract}
\noindent
Hybrid Monte Carlo and molecular dynamics simulations were used to investigate the interaction of light interstitials in multi-element Ni-based alloys. We show that light interstitials such as boron and oxygen fundamentally alter interfacial chemistry by reshaping alloy-element distribution and segregation. Oxygen adsorption drove boron migration from the grain boundary to the free surface, where it co-enriched with Cr, Fe, and Mo and formed BO3 trigonal motifs embedded within mixed-metal oxide networks. Oxygen also promoted M--O--M chain formation, including Nb2O5 clusters at the free surface. In the absence of oxygen, boron segregated to the grain boundary, altering local metal chemistry and underscoring a dynamic, environment-sensitive behavior. Following chlorine exposure, the oxidized surfaces retained strong O-mediated connectivity while forming new Cl--M associations, particularly with Nb and Cr, and exhibited further surface enrichment in Cr, Fe, and Mo. High-temperature MD simulations revealed a dynamic tug-of-war: chlorine exerted upward pull and disrupted weakly anchored sites, while Nb- and BO3-rich oxide motifs resisted deformation. A new stabilization mechanism was identified in which subsurface boron atoms anchored overlying Cr centers, suppressing their mobility and mitigating chlorine-driven displacement. These results demonstrate boron’s dual role as a modifier of alloy-element segregation and a stabilizer of oxide networks, and identify Nb as a key element in reinforcing cohesion under halogen attack. More broadly, this study highlights the need to track light interstitial cross-talk and solute migration under reactive conditions, offering atomistic criteria for designing corrosion-resistant surface chemistries in Ni-based superalloys exposed to halogenated or oxidative environments.
\end{abstract}

\section{Introduction}
Ni-based superalloys are critical structural materials in high-temperature environments such as gas turbine engines, nuclear reactors, and heat exchangers due to their exceptional mechanical strength, phase stability, and corrosion resistance. A key contributor to their oxidation resistance is the formation of protective surface oxide scales, such as Cr$_2$O$_3$, which serve as diffusion barriers to oxygen ingress and help limit further degradation under oxidizing conditions. However, as alloy chemistries become more complex and service conditions more aggressive, traditional understandings of oxidation behavior must be revisited. 

Park et al.~\cite{parkEffectsCrMo2019} found that in Ni--Cr--Mo--W--Al--Ti--Ta alloys, Cr$_2$O$_3$ offered robust protection at 850~$^\circ$C but became unstable at 1000~$^\circ$C, shifting the oxidation resistance to Al$_2$O$_3$. The role of refractory elements like Mo and W was also shown to be complex, influencing oxygen diffusivity, oxide porosity, and scale continuity. Similarly, Lapington et al.~\cite{lapingtonCharacterizationOxidationMechanisms2021} demonstrated that the oxidation kinetics in Cr$_2$O$_3$-forming Ni alloys were not governed by bulk Ti content, but rather by Ti and Nb segregation to oxide grain boundaries, underscoring the importance of short-circuit diffusion pathways in scale evolution. Long-term oxidation studies further illustrate the role of microstructure and grain boundary chemistry. Malacarne et al.~\cite{malacarneLongtermIsothermalOxidation2021} showed that Inconel 718 formed Cr$_2$O$_3$ externally and Al$_2$O$_3$ along grain boundaries, with extremely low oxidation rates at 700~$^\circ$C. In complementary work, Sanviemvongsak et al.~\cite{sanviemvongsakHighTemperatureOxidation2018} reported that continuous Cr$_2$O$_3$ layers formed only in Ni--Cr alloys with 15~wt\% Cr or more, while lower-Cr alloys exhibited discontinuous and porous NiO-rich scales. The stability of Cr$_2$O$_3$ was also found to be strongly dependent on oxygen partial pressure, with high pO$_2$ driving spinel formation (NiCr$_2$O$_4$) instead of stable Cr$_2$O$_3$. Additionally, Ye et al.~\cite{yeInfluenceNbAddition2021} showed that small additions of Nb to a Ni-based superalloy promoted compact oxide growth by stabilizing a TiO$_2$ overlayer and suppressing inward diffusion of oxygen and nitrogen. Going beyond the lab, Pedrazzini et al.~\cite{pedrazziniInServiceOxidationMicrostructural2018} examined Inconel 625 exhaust components under aggressive service conditions and observed grain boundary Cr depletion, Nb enrichment, and Mo-Si-rich precipitate formation, indicating a shift in interfacial chemistry during extended oxidation. These studies collectively point to the critical role of localized segregation and dopant-stabilized interfaces in controlling oxidation behavior. Supporting these experimental works, recent first-principles studies have simulated oxygen adsorption on compositionally complex alloys, evaluating adsorption energies, thermodynamic stability, and preferred surface chemical ordering~\cite{osei-agyemangSurfaceOxidationMechanism2019a, dolezalAdsorptionOxygenHigh2022b, khalidOxygenAdsorptionAbsorption2024, boakyeFirstprinciplesStudyOxide2025}. Most notably, these studies consistently report enhanced oxidation tendencies of refractory elements, highlighting their role in early-stage surface reactivity and scale development.

To highlight the role of light interstitials such as boron and carbon, Sudbrack et al.~\cite{sudbrackCharacterizationGrainBoundaries2016} analyzed grain boundary chemistry and minor phase evolution in the Ni-based disk alloy ME3 following long-term exposure at 815~$^\circ$C. They revealed that oxidation-driven chromium depletion at the grain boundaries correlated with dissolution of Cr-rich M$_{23}$C$_6$ carbides and subsequent nucleation of $\sigma$-phase in Cr- and Mo-enriched regions. Importantly, Mo- and Cr-rich M$_3$B$_2$ borides remained stable over 2,000-hour exposures, and boron segregation to the grain boundaries persisted across all conditions. These findings highlight the interconnected nature of oxidation behavior, elemental segregation, and dopant stability in determining grain boundary phase equilibria, motivating further investigation into how light elements like B and C influence near-surface chemistry in oxidizing environments. To extend understanding beyond experimental observations, Sushko et al.~\cite{sushkoMultiscaleModelMetal2015} developed a multiscale model for oxidation at grain boundaries in Ni--Cr and Ni--Al binary alloys. The study revealed key differences in oxidation mechanisms: Cr-selective oxidation led to porous Cr$_2$O$_3$ channels along grain boundaries in Ni--5Cr, while Ni and Al co-oxidation in Ni--4Al generated a mixed spinel oxide with more isotropic porosity highlighting the role of local chemistry and grain boundary structure in directing intergranular oxidation dynamics.

While protective oxide scales can significantly enhance corrosion resistance in oxygen-rich environments, they are far less effective under chlorine-containing atmospheres. Chlorine promotes an ``active oxidation'' mechanism wherein protective oxides such as Cr$_2$O$_3$ are converted into volatile metal chlorides, which evaporate or reoxidize away from the surface, continuously depleting the alloy of key elements. Wang et al.~\cite{wangChlorineinducedHightemperatureCorrosion2023} reviewed the behavior of heat-resistant alloys in chlorine-rich atmospheres, highlighting that grain boundaries and surface defects act as preferred sites for chlorine ingress and metal egress, further accelerating localized degradation. In multicomponent systems, the susceptibility to chlorine attack is highly dependent on element-specific affinities for chlorine and oxygen, which shape both the composition and integrity of the resulting oxide scale. Experimental work by Chen et al.~\cite{chenHotChlorinationCorrosion2020} reinforces the catalytic role of alkali halides in chlorine-driven degradation. They demonstrated that NaCl reacts with NiCl$_2$ to form a eutectic liquid phase, NiNa$_{0.33}$Cl$_{2.33}$, which facilitates internal chlorine transport and suppresses chloride volatilization. Corrosion was most severe at 700~$^\circ$C, where nearly 97\% of the nickel was converted to chlorides under optimal NaCl loading. Alloying additions have also been shown to mediate chlorine attack. Wang et al.~\cite{wangSignificantRolesNb2020} found that Nb and Mo additions to ferritic stainless steels promoted the formation of Laves phases and dense oxide scales, both of which inhibited oxygen ingress and reduced Cr depletion.

Work on Fe addition was performed by Xie et al.~\cite{xieEffectFeCorrosion2019}, who reported that Fe additions to Ni--30Cr alloys enhanced corrosion resistance by promoting continuous Cr$_2$O$_3$ bands and altered internal oxidation zone morphology. Iron also played a role in intergranular carbide formation, which may influence long-term degradation behavior. More recently, Louren\c{c}o et al.~\cite{lourencoInfluenceIronContent2021} studied the electrochemical response of Inconel 625 alloys with varying Fe content and showed that low Fe additions promoted the formation of Nb-rich monocarbide precipitates, while higher Fe contents suppressed monocarbide formation. They also revealed that monocarbide phases promoted localized corrosion, and the absence of monocarbide in high-Fe alloys reduced potential gradients and improved corrosion resistance. These findings suggest that Fe, like other minor alloying elements, influenced corrosion not only through bulk composition, but also by directing microstructural evolution and interfacial electrochemical stability.

Similarly, Nb has been shown to play a complex role in oxidation and interfacial chemistry. Although typically a minor constituent in Ni-based alloys, Nb consistently enriches at both free surfaces and internal interfaces during oxidation, forming distinct Nb$_2$O$_5$ and NbC-rich domains~\cite{haInvestigatingOxideFormation2024, gengOxidationBehaviorAlloy2007, millerMechanismOxygenEnhanced2001}. Experimental studies have identified Cr- and Nb-rich bilayer oxide scales~\cite{haInvestigatingOxideFormation2024}, Nb-enriched Cr$_2$O$_3$ outer layers~\cite{gengOxidationBehaviorAlloy2007}, and co-segregation of Nb with Mo and O at grain boundaries and intercellular regions~\cite{staronQuantitativeMicrostructuralCharacterization2022}. These behaviors stem from Nb’s strong oxygen affinity and its ability to alter the local thermodynamic environment, stabilizing certain oxide or carbide phases while impeding oxygen and nitrogen ingress~\cite{yeInfluenceNbAddition2021}. As such, Nb’s role in oxidation resistance is both direct, through the formation of protective oxide clusters, and indirect, through the stabilization or disruption of neighboring interfacial species.

Like Fe and Nb, boron is a minor alloying addition known to influence interfacial stability, though its oxidation behavior in alloy systems has received less attention. While boron is widely recognized for its role in grain boundary strengthening and phase stabilization in Ni-based superalloys, its influence on surface oxidation behavior remains underexplored. Several experimental studies have shown that boron can improve oxidation resistance by altering near-surface chemistry and promoting the formation of protective boride phases. Peruzzo et al.~\cite{peruzzoHightemperatureOxidationSintered2017} observed reduced oxidation rates and enhanced scale adhesion in boron-doped stainless steels, with EDS mapping revealing co-segregation of boron and chromium at the oxidized surface. Similarly, Lee et al.~\cite{leeEffectBoronCorrosion2007a} and Zhang et al.~\cite{zhangEvolutionMicrostructureProperties2016a} reported the formation of (Cr,Fe)-borides at the surface of oxidized boron-doped high-entropy alloys, which were associated with improved corrosion resistance and reduced surface degradation.

Together, these studies underscore the sensitivity of oxidation and degradation behavior to both alloy composition and microstructural features, particularly at grain boundaries and free surfaces. While considerable experimental work has highlighted the roles of substitutional and interstitial elements in mediating oxidation resistance and corrosion susceptibility, a detailed understanding of how these effects emerge from local chemical ordering and segregation remains incomplete. In this study, we investigate the oxidation and chlorine-induced degradation behavior of a chemically complex Ni-based alloy (Ni--Cr--Fe--Mo--Nb--Si) using a hybrid Monte Carlo and molecular dynamics simulation approach. The simulation cell includes both a free surface and a grain boundary, offering the ability to resolve dopant partitioning and chemical ordering across competing interfacial environments. We compare undoped and boron-doped systems to assess how light interstitials alter near-surface chemical ordering, oxygen adsorption, and subsequent chlorine reactivity. Importantly, the introduction of co-interstitial interactions (e.g., B + O, B + O + Cl), new MC moves enabling direct B--O--Cl competition, and detailed analysis of interface-specific enrichment and degradation establish this work as a distinct and novel extension beyond prior studies of light interstitial segregation in bulk and GB-only systems. By simulating both oxygen adsorption and subsequent chlorine exposure, this work provides atomistically resolved insight into how dopants shape the surface and grain boundary chemistry that governs early-stage degradation mechanisms in Ni-based superalloys.

\section{Methods}
\begin{figure}[H]
    \centering
    \includegraphics[width=\linewidth]{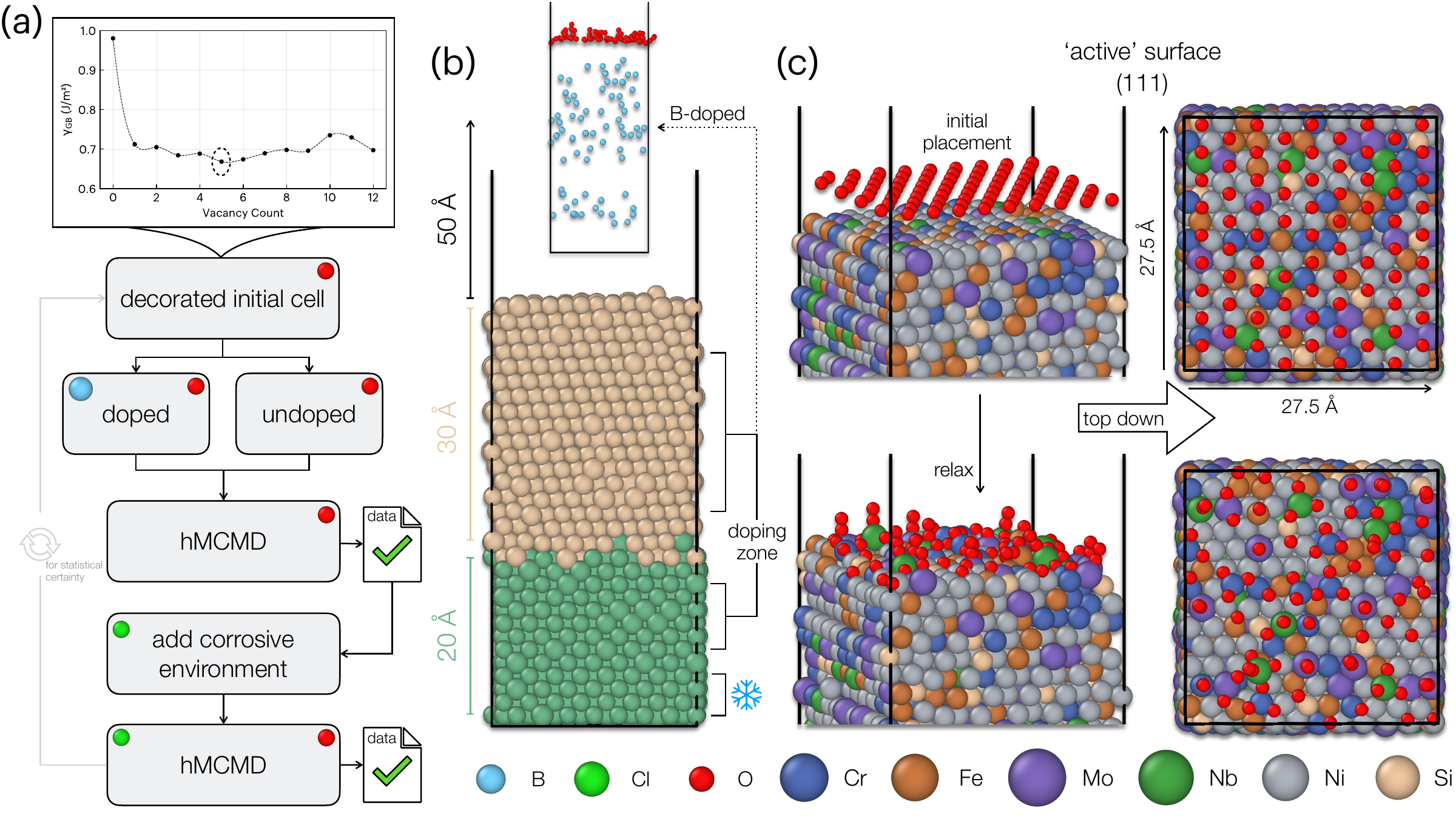}
    \caption{(a) Step-by-step outline of the computational routine. The grain boundary energy ($\gamma_{\rm GB}$) is plotted as a function of vacancy count, with the selected structure highlighted by a dashed oval. (b) A view of the selected system, prepared for the surface study, with the grain segmentation modifier turned on. The snow flake marks the frozen layers. (c) A closer look at the `active' (111) surface viewed side-on and top-down to show the initial oxygen placement (top) and relaxed oxygen placement (bottom). These visualizations are as seen in OVITO.}
    \label{fig:overview}
\end{figure}
\subsection{Preparing the Surface Structures}
The surface structure, visualized in OVITO \cite{stukowskiVisualizationAnalysisAtomistic2009a}, is shown in Fig. \ref{fig:overview} from multiple perspectives. The simulation cell was initialized as a grain boundary (GB) structure of pure Ni (Fm$\bar{3}$m cubic crystal system) using Atomsk \cite{hirelAtomskToolManipulating2015}. The top grain (color-coded gold) was oriented such that the (111) surface was normal to the $z$-axis ($\hat{g}_{111} \parallel \hat{z}$). The bottom grain (color-coded green) was oriented such that the $x$-, $y$-, and $z$-axes aligned with the (100), (010), and (001) crystallographic directions, respectively. To allow free surface and GB relaxation, a 50~\AA\ vacuum was placed along the $\hat{z}$ direction. To optimize the GB configuration, a systematic vacancy-driven relaxation was performed: one Ni atom was iteratively deleted from the GB core, followed by 5,000 steps of a canonical ($NVT$) molecular dynamics (MD) simulation at 1000 K and subsequent conjugate gradient (CG) relaxation (more details provided in Sec. \ref{sec:comp_details}). The structures were recorded as a function of atomic deletions, which continued until the grain boundary energy ($\gamma_{\rm GB}$) converged (further details are provided in Supplemental Materials). From the stabilized region, a representative structure was selected for further analysis, as indicated by the dashed oval in Fig.~\ref{fig:overview}a.

The selected structure consisted of 30 atomic layers along the $z$-axis, with the bottom four layers frozen to emulate a bulk substrate. These layers were held fixed in structural relaxation and ignored during the Monte Carlo routine, ensuring that the substrate atoms remained unchanged during the simulations. The GB was positioned approximately seven layers above the substrate layers and 18 layers below the free surface. The simulation system consisted of 4,087 atoms and was randomly decorated according to the following chemical composition (in at\%): Cr\textsubscript{15}, Fe\textsubscript{15}, Mo\textsubscript{10}, Nb\textsubscript{5}, Ni\textsubscript{50}, and Si\textsubscript{5}. This setup enabled the simultaneous investigation of atomic ordering and chemical distribution under the competing influences of a free surface and GB. Furthermore, simulations were performed with and without the inclusion of the light interstitial boron in the bulk to assess its effects on segregation tendencies, atomic ordering, and structural stability. Interstitial dopants were carefully introduced into the interstitial medium, avoiding both the frozen layers and the grain boundary or free surface, collectively referred to as the `doping zone' in Fig.~\ref{fig:overview}b. This prevented artificial bias, enabling the MC routine to naturally explore boron’s segregation behavior.

Because oxidation of the environment-facing surface is a critical factor in high-temperature applications, the free surface was exposed to an oxygen-rich environment and equilibrated in both undoped and interstitial-doped systems. Oxidation was modeled by introducing a sheet of 81 oxygen atoms above the (111) active surface. The lateral spacing between oxygen atoms in the layer was set to 1.52~\AA, and their vertical distance from the surface was determined using van der Waals radii ($r_W$)~\cite{bondiVanWaalsVolumes1964}. Specifically, the placement height followed the criterion defined in Eq.~\ref{eq:surface_placement}:

\begin{equation}\label{eq:surface_placement}
d = \frac{1}{2}\big(r_{\mathrm{O}} + \max(r_{j})\big)
\text{,}
\end{equation}

where $j$ represents the metallic species present in the first surface layer. This ensured appropriate spacing between the oxygen layer and the variably sized surface atoms. For clarity, the resulting systems are hereafter referred to as ``\clean'' (undoped, no oxygen), ``\Bonly'' (boron-doped, no oxygen), ``\Oonly'' (undoped, with oxygen), and ``\BO'' (boron-doped, with oxygen), consistent with the labeling used in all figures. To simulate environmental degradation under more aggressive conditions, a second round of simulations was conducted on the previously oxidized and equilibrated systems (\Oonly\ and \BO) by introducing a layer of chlorine atoms (49 Cl atoms), positioned using the same spacing rules as for oxygen. Separate MC simulations were then performed in the presence of both oxygen and chlorine to emulate Cl-assisted degradation and assess how the presence of boron influences chemical reactivity of the oxidized surface and resistance to chlorine-induced volatilization. This sequential exposure approach reflects realistic service environments, where high-temperature alloys first form stable oxide layers before encountering chlorine-containing atmospheres. Modeling chlorine interaction with this chemically evolved surface enables a more accurate representation of long-term degradation behavior and the role of dopants in stabilizing complex oxidized interfaces.

\subsection{Computational Details}\label{sec:comp_details}
Monte Carlo (MC) simulations were performed on both undoped and boron-doped systems to explore the equilibrium distribution of light interstitial elements in the presence of a free surface and grain boundary. Each system was subjected to two independent MC runs to ensure statistical robustness and capture potential configurational variability. Additional runs were considered; however, the observed consistency between trials indicated that further sampling would yield diminishing returns. These simulations were conducted both prior to and following oxygen adsorption, enabling direct comparison of dopant segregation behavior in clean and oxidized environments. The resulting ensemble of configurations provides insight into the thermodynamic driving forces governing interstitial segregation and surface reactivity under high-temperature conditions. Simulations were run until thorough equilibration was achieved in both the total energy and the short-range ordering of the systems. The MC algorithm trialed the following moves:
\begin{enumerate}
    \item \textbf{Swap atomic positions of metallic constituents:} Select two metallic atoms of different chemical types and attempt to swap their positions within the lattice.
    \item \textbf{Relocate a light interstitial atom near a new metallic host:} Select a light interstitial atom and attempt to place it near a different metallic atom within its nearest neighbor shell.
    \item \textbf{Introduce proximity between two light interstitial atoms:} Select two light interstitial atoms and attempt to place one of them within the first nearest neighbor shell of the other. If the first shell is fully occupied, the placement is attempted in the second nearest neighbor shell.
    \item \textbf{Separate two neighboring light interstitial atoms:} If a pair of neighboring light interstitials exists, select one and attempt to move it away from its nearest interstitial neighbor. This move balances the proximity adjustment, ensuring an opportunity to promote both segregation and aggregation of light interstitials.
    \item \textbf{Swap a metal neighbor of a light interstitial atom:} For a selected light interstitial atom, identify one of its nearest metallic neighbors and swap its position with that of another metal atom of a different chemical type.
    \item \textbf{Swap two dissimilar light interstitial atoms:} If two distinct light interstitial species are present at the free surface, randomly select a dissimilar pair within the nearest-neighbor shell and attempt to exchange their positions. This surface-restricted move facilitates local chemical rearrangements and allows competition between interstitial species for favorable binding environments.

\end{enumerate}
These moves were designed to explore the configurational space of the system effectively, balancing local aggregation and dispersion tendencies of the light interstitial atoms within the metallic matrix and at the free surface. Please note, adsorbed O and Cl atoms were treated as light interstitials, allowing them to participate in MC moves. Nearest neighbors were located within a cutoff radius of 2.75~\AA. The move was accepted or rejected based on the potential energy difference between the final and initial states, following the Metropolis criterion \cite{metropolisEquationStateCalculations1953}. The probability of acceptance is expressed as a Boltzmann probability:
\begin{equation}\label{eq:metropolis}
    P(\Delta U) = \min\{1, e^{-\Delta U/{k_{\rm B} T_{\rm sim}}}\}\rm ,
\end{equation}
where $\Delta U$ is the potential energy difference between the final and initial states, $k_\mathrm{B}$ is the Boltzmann constant in eV/K, and $T_{\rm sim}$ is the simulation temperature in K, which was set to 1073 K (800 \textsuperscript{o}C) for this work. The potential energy difference for each trial was calculated using the Large-scale Atomic/Molecular Massively Parallel Simulator (LAMMPS) software \cite{thompsonLAMMPSFlexibleSimulation2022a} and version 5.0.0 of the universal core neural network Preferred Potential (PFP) \cite{takamotoUniversalNeuralNetwork2022} with the D3 correction implemented through Matlantis \cite{Matlantis}. Structural relaxation was performed using CG minimization with an energy convergence criterion of $1 \times 10^{-6}$~eV and a force convergence criterion of 0.01~eV/\AA. The accuracy and transferability of the PFP framework have been extensively validated by its developers. In our prior work, we further benchmarked PFP predictions against density functional theory (DFT) for Ni-based systems containing light interstitials such as B, C, H, and N~\cite{DOLEZAL2025121221}. For completeness, a summary of key DFT-to-PFP validation results are provided in Table~S2 and Table~S3 of the Supplemental Materials. In addition to these benchmarks, we note that many of the trends and mechanistic insights uncovered in this study show strong agreement with experimental observations reported in the literature. These points of corroboration, highlighted throughout the Results and Discussion section, provide further confidence in the reliability and physical relevance of our simulation framework.

All MC-equilibrated structures were subsequently relaxed using a 20~ps MD simulation in the $NVT$ ensemble at 1073~K. To assess the reproducibility and temporal stability of early-stage dynamics, one representative configuration of the \BO\ + Cl system was extended to 180~ps. Snapshots recorded every 20~ps revealed minimal variation in surface SRO and connectivity metrics over time, confirming statistical equilibration of the key interfacial behaviors. Based on this assessment, the 20~ps duration was used for all production simulations. This timescale captured meaningful interactions between adsorbed chlorine and the evolving oxide network, including surface restructuring, bond-angle fluctuations, and transient metal--chlorine complexes.

\subsection{Analysis Methods}

\subsubsection*{Chemical Ordering}

The Warren-Cowley short-range order (SRO) parameter ($\alpha_{ij}$) \cite{cowleyApproximateTheoryOrder1950, cowleyShortLongRangeOrder1960} was examined to quantify the degree of atomic ordering in the system and was defined as:
\begin{equation}
    \alpha_{ij}(r) = 1 - \frac{P^{*}_{ij}(r)}{c_j} \, ,
\end{equation}
where the probability term $ P^{*}_{ij}(r) $ depends on the local environment:
\begin{equation}
P^{*}_{ij}(r) = 
\begin{cases}
    P_{ij}(r) = \frac{N_{ij}(r)}{N_i(r)} & \text{if atom } i \text{ is at the free surface}, \\
    \tilde{P}_{ij}(r) = P_{ij}(r) \cdot \left( \frac{N_i(r)}{N^{\text{expected}}_i} \right) & \text{if atom } i \text{ is in the bulk or GB}.
\end{cases}
\end{equation}
Here, $ N_{ij}(r) $ is the number of neighbors of species $ j $ around atom $ i $ within a cutoff radius $ r $, $ N_i(r) $ is the total number of neighbors around atom $ i $, and $ N^{\text{expected}}_i $ is the ideal coordination number based on the site type of atom $ i $ (e.g., 12 for FCC sites, 6 or 4 for octahedral or tetrahedral interstitials, respectively). This piecewise formulation ensures that coordination losses intrinsic to surfaces are treated differently than coordination deviations near defects, preserving physical meaning in chemically or structurally heterogeneous environments. Additionally, to maintain physical consistency across chemically diverse systems, the reference concentration $ c_j $ was computed locally within each analysis region and normalized against the number of metal atoms present (excluding light elements such as O, B, and Cl). That is, for SRO calculations focused on the free surface or the bulk + GB region, $ c_j $ reflects the concentration of species $ j $ within that specific region, divided by the total number of metal atoms in the same region. This approach ensures a standardized denominator across all systems, regardless of the presence or absence of oxygen and chlorine, and prevents artificial distortions in $ \alpha_{ij}(r) $ due to global averaging in structurally or chemically heterogeneous environments. The results for $\alpha_{ij}$ were averaged over the two independent MC simulations for each system to improve statistical reliability.

Additionally, the atomic composition per atomic layer was calculated as a function of $z$. For each atomic layer $z$, the fractional concentration of species $i$ was defined as

\begin{equation}\label{eq:z_scan}
x^{(i)}_z = \frac{N^{(i)}_z}{\sum_{j \in \mathcal{M}} N^{(j)}_z} \quad \text{for metals}, \qquad
x^{(i)}_z = \frac{N^{(i)}_z}{\sum_{j \in \mathcal{I}} N^{(j)}_z} \quad \text{for interstitials}
\end{equation}
where $N^{(i)}_z$ is the number of atoms of species $i$ in atomic layer $z$, and the summations are restricted to subsets of species: $\mathcal{M}$ represents the set of all metallic elements in the simulation, while $\mathcal{I}$ denotes the set of interstitial species (e.g., B, Cl, and O). The notation $j \in \mathcal{M}$ and $j \in \mathcal{I}$ explicitly filters the denominator to include only metals or only interstitials, respectively. This formulation distinguishes between metal and interstitial populations, and clarifies why the total composition in a layer does not sum to 100\% when metals and interstitials are plotted together. This provides a detailed, normalized view of chemical distributions near the grain boundary and free surface. 

To quantify elemental enrichment or depletion at interfaces, atomic fractions were computed and grouped by either the free surface region or the grain boundary zone. For each species $i$, the local atomic fraction within the target region was compared to its full atomic fraction in the system. An enrichment metric was defined as the difference:
\begin{equation}
\Delta^{(i)}_{\text{zone}} = x^{(i)}_{\text{zone}} - x^{(i)}
\end{equation}
where $x^{(i)}_{\text{zone}}$ and $x^{(i)}$ represent the atomic fractions of species $i$ in the selected interfacial zone and throughout the complete system, respectively. Positive values of $\Delta^{(i)}_{\text{zone}}$ indicate enrichment at the interface, while negative values denote depletion. This direct difference metric enables intuitive visualization of segregation tendencies in heatmap form across multiple systems and species.

\subsubsection*{Free Energy Analysis of Surface Oxidation and Degradation}

To evaluate the thermodynamic stability of surface oxidation across doped and undoped systems, the Gibbs free energy of oxidation ($\Delta G_{\text{O}}$) was computed as a function of oxygen partial pressure, $p_{\text{O}_2}$, at 1073 K. This analysis was motivated by the MC simulation results which revealed strong segregation of boron to the free surface in the presence of oxygen. There are three key MC-equilibrated configurations: (i) the \clean\ reference system, (ii) the \Bonly\ system, and (iii) the \Oonly\ and \BO\ systems. For each configuration, $\Delta G_{\text{O}}$ was computed relative to the \clean\ alloy baseline as:
\begin{equation}
\Delta G_{\text{O}}(T, p_{\text{O}_2}) = \frac{1}{n_{\rm O}}\big(U_{\text{oxidized}}(T) - U_{\text{clean}}(T) - n_{\text{O}} \mu_{\text{O}}(T, p_{\text{O}_2})\big),
\end{equation}
where $U_{\text{oxidized}}(T)$ and $U_{\text{clean}}(T)$ are the average internal energies obtained from the final $NVT$ MD simulations at 1073 K for the oxidized and clean systems, either undoped or boron-doped, depending on the comparison. The term $n_{\text{O}}$ is the number of oxygen atoms added to the system, and $\mu_{\text{O}}(T, p_{\text{O}_2})$ is the temperature- and pressure-dependent chemical potential of oxygen gas. This approach allows a consistent evaluation of oxidation thermodynamics by isolating the chemical effects of oxygen adsorption. The oxygen chemical potential $\mu_{\text{O}}(T, p_{\text{O}_2})$ was evaluated as:
\begin{equation}
\mu_{\text{O}}(T, p_{\text{O}_2}) = \mu_{\text{O}}^\circ(T) + \tfrac{1}{2} k_{\text{B}} T \ln\left(\frac{p_{\text{O}_2}}{p^\circ}\right),
\end{equation}
where $\mu_{\text{O}}^\circ(T)$ is the standard-state chemical potential of oxygen at 1073 K, evaluated as $-1.11$ eV per O atom based on interpolation from the JANAF thermochemical tables~\cite{janafO2}, and $p^\circ$ is the standard pressure (1~atm). This formulation captures the dependence of oxidation driving force on environmental conditions, enabling direct comparison between the undoped and boron-doped systems under realistic high-temperature service conditions. The construction of $\Delta G_{\text{O}}$ vs.\ $\log p_{\text{O}_2}$ curves provides a framework to evaluate how light interstitial dopants influence the thermodynamic stability of surface oxidation across chemically complex alloy systems. In this work, $\Delta G_{\text{O}}$ was computed relative to each system's MC-equilibrated clean structure, and thus reflects the oxidation thermodynamics within a given chemical environment. While this allows for direct comparison of oxidation trends across configurations of the same composition, absolute comparisons between boron-doped and undoped systems require a shared reference state and are not made here.

To evaluate the susceptibility of undoped and boro-oxidized surfaces to chlorine-induced degradation, the Gibbs free energy change associated with chlorine adsorption, $\Delta G_{\text{Cl}}$, was computed using an analogous approach to that described for oxygen. In this case, Cl-exposed configurations (\Oonly\ and \BO) were compared against their pre-chlorinated, oxidized baselines, with

\begin{equation}
\Delta G_{\text{Cl}}(T, p_{\text{Cl}_2}) = \frac{1}{n_{\rm Cl}}\big(U_{\text{Cl}}(T) - U_{\text{O}}(T) - n_{\text{Cl}} \mu_{\text{Cl}}(T, p_{\text{Cl}_2})\big),
\end{equation}

where $U_{\text{Cl}}(T)$ and $U_{\text{O}}(T)$ are the internal energies of the Cl-exposed and pre-chlorinated oxidized systems, respectively. The chemical potential of chlorine gas, $\mu_{\text{Cl}}(T, p_{\text{Cl}_2})$, was evaluated using the same ideal gas expression as for oxygen, with $\mu_{\text{Cl}}^\circ(T)$ taken as $-1.26$ eV per chlorine atom at 1073 K based on JANAF data~\cite{chasem.w.jr.NISTJANAFThermochemicalTables1982}. This construction provides a consistent thermodynamic framework to assess chlorine adsorption within each system. Because $\Delta G_{\text{Cl}}$ is referenced to the MC-equilibrated oxidized structure for each composition, comparisons are valid only between configurations of the same chemical environment. While a more negative $\Delta G_{\text{Cl}}$ in the \BO\ + Cl system may reflect enhanced stabilization of chlorine due to boron incorporation and its influence on local chemical ordering, this does not directly imply greater susceptibility to chlorine-induced degradation. The implications depend on subsequent degradation behavior and the chemical nature of the resulting interfacial species. To contextualize the thermodynamic trends, the final 20 ps $NVT$ MD simulation provides an opportunity to observe the dynamic response of the metal-oxide network.

\section{Results and Discussion}
\begin{figure}[H]
    \centering
    \includegraphics[width=\linewidth]{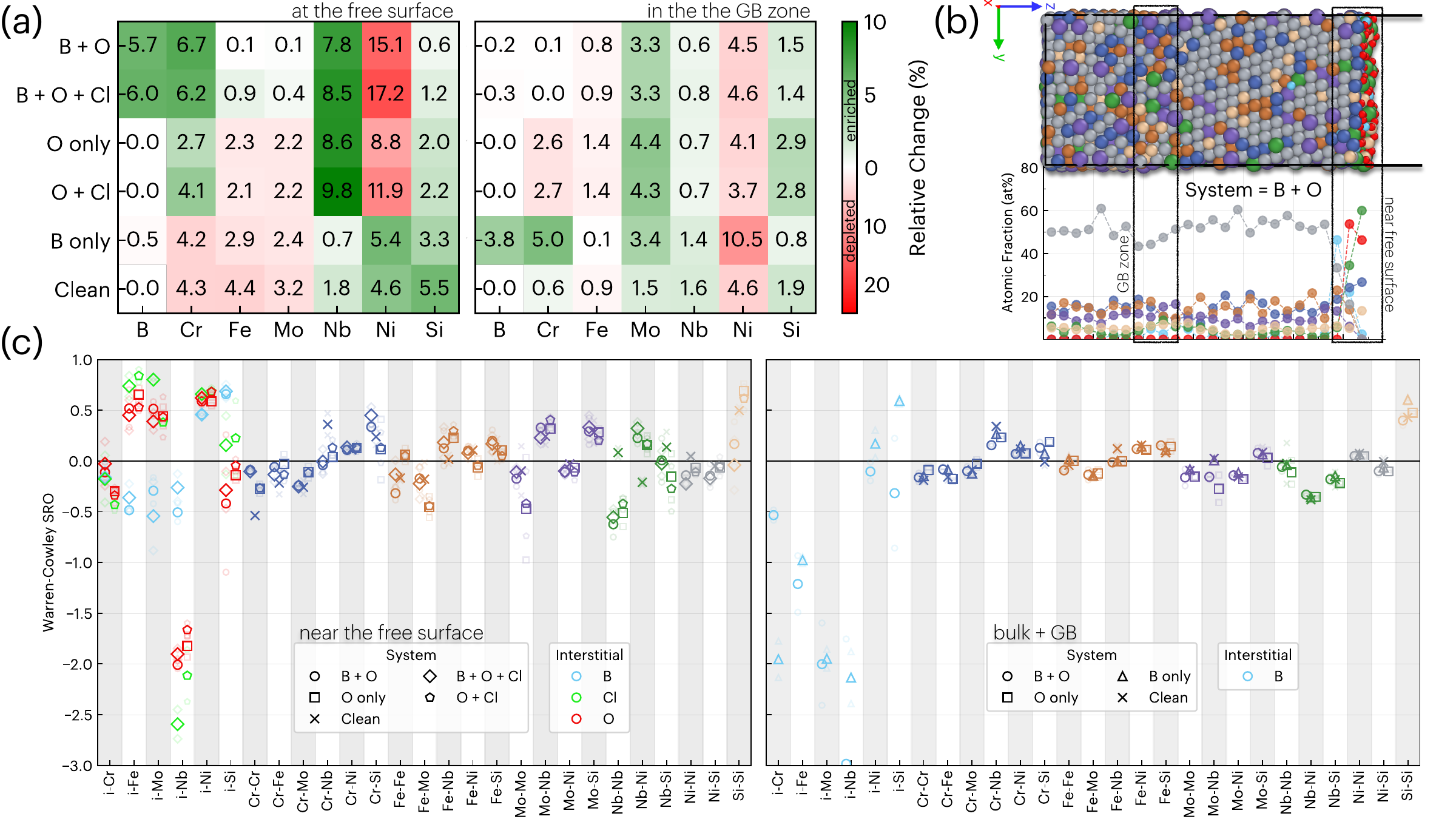}
    \caption{(a) Relative change in the atomic percent for each element at the free surface (left) and grain boundary (GB) zone (right) compared to its total atomic percent. A value of zero means there was no change, positive values indicate enrichment (green), and negative values indicate depletion (red). Values of 0.0 are assigned to boron in systems where it was not present. (b) Layer-by-layer elemental distribution as a function of the $z$-axis, illustrating spatially resolved segregation behavior across the simulation cell. (c) Warren-Cowley short-range order (SRO) parameters for interstitial-metal (i-M) and metal-metal (M-M') pairs, calculated separately near the free surface (left) and within the GB + bulk region (right). Different markers distinguish each system, while metal-metal and interstitial-metal pairs are color-coded based on the identity of the metal or interstitial species involved (B--M pairs are light blue, Cr-M pairs or blue, Fe-M pairs are orange, etc.). Semi-transparent markers of the same type indicate individual values (run 1 and run 2), while the corresponding average is overlaid in full color. A small horizontal offset is applied to prevent marker overlap and improve visual clarity.}
\label{fig:ordering}
\end{figure}

Many of the key behaviors: segregation trends, oxide formation, and chlorine-induced chemical changes, are interrelated and are most clearly resolved when plotted together in a single comparative figure, Fig. \ref{fig:ordering}. For clarity, however, the discussion is organized in two parts: first, the chemical and structural evolution of each system upon oxygen adsorption is examined; second, the response of the oxidized systems (\BO\ and \Oonly) to subsequent chlorine exposure is analyzed. This enables a more focused interpretation of how each light interstitial alters surface chemistry under sequential environmental challenges.
\subsection*{Chemical Ordering and Surface Segregation During Oxidation}
\subsubsection*{Free Surface Chemical Enrichment and Oxide Formation}
\begin{figure}[H]
    \centering
    \includegraphics[width=\linewidth]{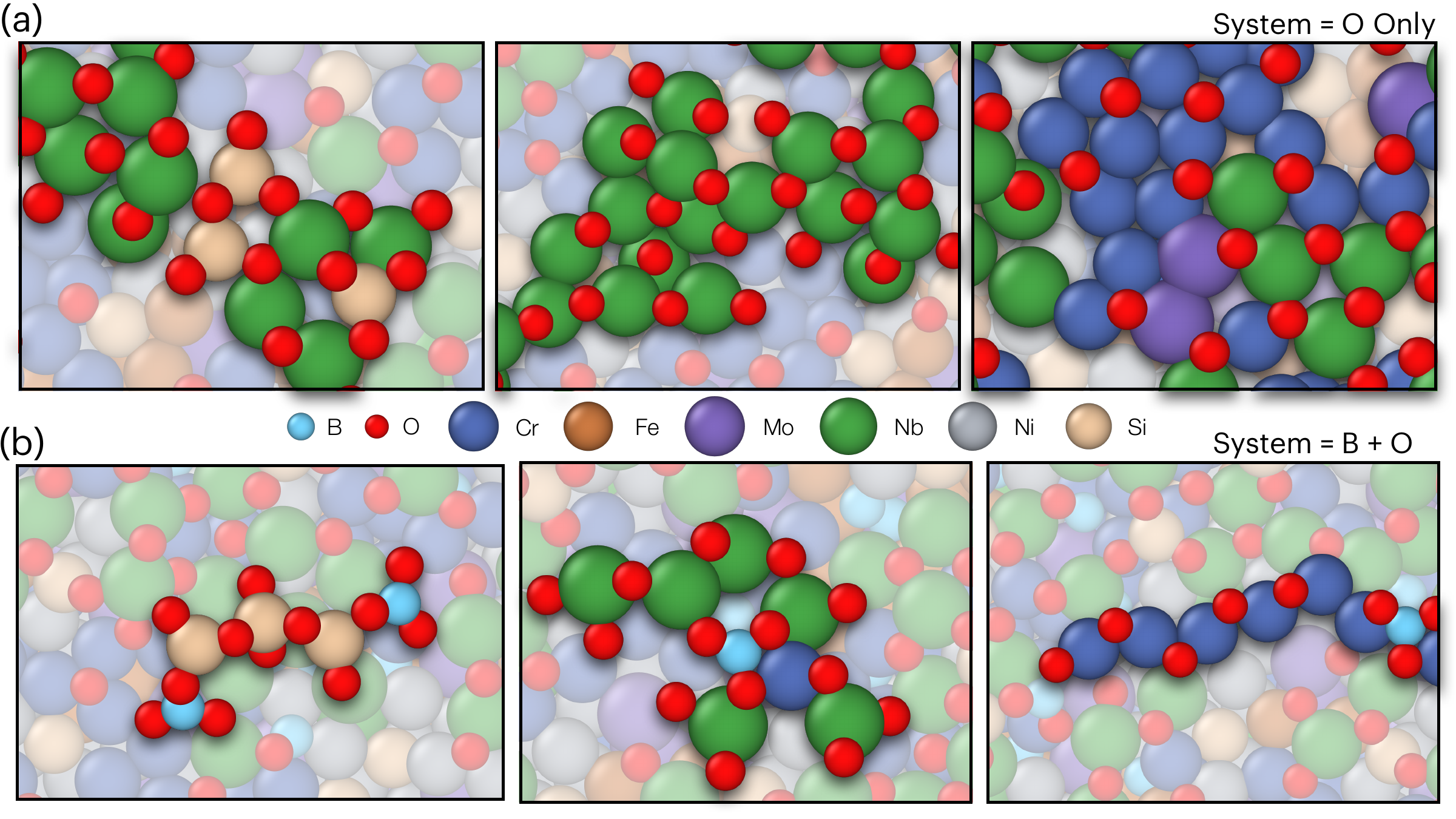}
    \caption{(a) Representative metal oxide clusters and chain-like motifs formed on the \Oonly\ free surface. (b) Corresponding structures from the \BO\ free surface, where BO$_3$ trigonal units are observed. These units consist of a boron atom centrally coordinated to three in-plane oxygen atoms, forming planar triangles embedded within both metal and mixed-metal oxide domains. The incorporation of boron promotes the development of structurally diverse and polymerized oxide networks. Each subpanel depicts a different surface region to illustrate the variety of oxygen coordination environments, including M--O--M chains, isolated oxide units, and early-stage network connectivity.}
    \label{fig:oxide_motifs}
\end{figure}

Beginning with only oxygen present, the \Oonly\ system demonstrates that oxygen adsorption significantly alters the free surface composition, with Nb, Cr, and Si exhibiting notable enrichment. Mo and Fe also exhibit reduced depletion compared to the \clean\ surface, indicating a more favorable chemical environment in the presence of oxygen. These elements are well-known oxide formers, and their preferential localization near the surface is consistent with their strong chemical affinity for oxygen. This trend is corroborated by SRO values in Fig.~\ref{fig:ordering}c (red squares), where O--Nb and O--Cr pairs exhibit negative values, reflecting preferential coordination. The resulting oxide motifs, shown in Fig.~\ref{fig:oxide_motifs}a, consist primarily of Cr--O--Cr and Nb--O--Nb chains, along with localized Fe--O, Mo--O, and Si--O bonds embedded within the oxide network, forming a chemically diverse patchwork of early-stage oxide domains. A strong signal of Ni depletion is also observed at the surface, likely resulting from oxygen-induced chemical ordering. This is particularly notable given that Ni was enriched in the \clean\ system. Although both Nb and Si were present at 5~at\% in the mixed-metal system, only Nb exhibited pronounced surface enrichment. These findings suggest that Nb exhibits a stronger oxygen affinity and surface-segregation tendency than Si under the simulated conditions. Nb’s preferential segregation likely stems from its high oxygen affinity and capacity to form multivalent coordination environments, reinforcing its role in early-stage surface passivation.

The enrichment of Cr and Nb at the free surface, along with the emergence of Nb$_2$O$_5$ structural motifs, is well supported by experimental studies of Inconel oxidation. Ha et al.~\cite{haInvestigatingOxideFormation2024} observed that the oxide scale in Inconel 718 at high temperatures formed a Cr-rich upper layer and a Nb-rich lower layer composed of Cr$_2$O$_3$ and Nb$_2$O$_5$. Geng et al.~\cite{gengOxidationBehaviorAlloy2007} further identified a Nb-rich Cr$_2$O$_3$ outer scale on Inconel 718 after prolonged oxidation at 1100\,°C, reinforcing the interpretation that Nb is surface-active and oxidizes readily despite its low bulk concentration. These observations corroborate Nb’s thermodynamic tendency to accumulate at chemically active sites and interact with interstitial species. Ye et al.~\cite{yeInfluenceNbAddition2021} demonstrated that Nb addition enhances oxidation resistance not only through Nb--O bonding, but also by promoting the formation of a continuous CrTaO$_4$ interlayer beneath the Cr$_2$O$_3$ scale. This additional layer reduced spallation and inhibited both oxygen and nitrogen diffusion. Such findings emphasize that an individual element’s contribution extends beyond direct chemical interactions, altering the local thermodynamic environment in ways that stabilize neighboring phases and promote new modes of chemical ordering.

This surface reactivity is mirrored at internal interfaces, where Nb is also known to segregate and contribute to the formation of discrete oxide domains and carbide precipitates~\cite{millerMechanismOxygenEnhanced2001, smithRoleNiobiumWrought2005, zhaoPrecipitationStabilityMicroproperty2018, zhangSynergyPhaseMC2024}. Staro\'{n} et al.~\cite{staronQuantitativeMicrostructuralCharacterization2022} observed strong Nb segregation to intercellular regions in Inconel 625, where it frequently co-localizes with O and Mo, contributing to the formation of NbC and Laves phase precipitates at internal interfaces. The simulated formation of Nb$_2$O$_5$ oxide motifs is consistent with experimental observations of Nb$_2$O$_5$ enrichment at grain boundaries and crack tips under oxidizing conditions~\cite{millerMechanismOxygenEnhanced2001}. These findings suggest that the atomistic tendency for Nb to form ordered, multivalent oxide clusters is not limited to external surfaces, but also governs the evolution of chemically distinct, oxidation-sensitive interfacial regions throughout the alloy microstructure. This behavior is driven by Nb’s strong oxygen affinity, which underlies both its surface segregation and its participation in localized oxidation reactions. As such, Nb’s role should be explicitly considered in models of oxidation-induced degradation, particularly at structurally sensitive or defect-rich regions. Collectively, these findings validate the segregation and ordering behavior observed in the present work, from Nb and Cr enrichment to Ni depletion in oxygen-rich regions, and emphasize Nb’s critical role in early-stage oxidation and interfacial stabilization or destabilization, depending on its localization at external surfaces versus internal interfaces.

Comparison of the \BO\ and \Oonly\ enrichment and depletion values at the free surface in Fig.~\ref{fig:ordering}a shows that the addition of boron markedly alters surface composition. A strong preference for surface segregation in response to adsorbed oxygen is further demonstrated in Fig.~\ref{fig:segregation}, where boron accumulates at the free surface only when oxygen is present. In contrast, in the absence of surface oxygen, boron preferentially migrates toward the grain boundary region, indicating that its segregation behavior is strongly mediated by local chemical conditions. Consequently, boron significantly enhances the presence of Cr, Fe, and Mo at the free surface relative to the \Oonly\ system, accompanied by modest reductions of 1\% in Nb and 1.5\% in Si content. This redistribution is reflected in the negative SRO parameters, $\alpha_{\rm BM}$, for M = Cr, Fe, Mo, and Nb in Fig.~\ref{fig:ordering}c (light blue circles), indicating preferential chemical association between boron and these metallic species at the surface. The enhanced surface presence of Cr, Fe, and Mo suggests that boron not only segregates to the surface but also co-localizes with chemically compatible metals. This co-segregation behavior is likely thermodynamically favorable due to boron's strong affinity for forming borides or mixed oxide species with these elements.

The resulting local enrichment and chemical ordering with oxide-forming metals support the development of structurally robust and chemically diverse oxide networks. In particular, Cr--O, Mo--O, Nb--O, and Si--O motifs incorporate BO$_3$ units, as demonstrated in Fig.~\ref{fig:oxide_motifs}b, highlighting boron’s role in facilitating the formation of polymerized, multicomponent oxide domains. These motifs are known to enhance the protective qualities of surface oxide scales, indicating that boron plays a dual role: as a surface-active element and as a network-forming species that promotes passivation. During oxidation of boron coatings, boron has been shown to migrate toward the surface and form a distinct top-layer oxide, either as an amorphous B$_2$O$_3$ cap or as part of a boron-enriched passivation layer~\cite{glechnerInfluenceNonmetalSpecies2021, yunactiReplacingToxicHard2023}. These surface-localized boron oxides act as effective diffusion barriers, limiting the ingress of oxygen and chloride~\cite{yushkovElectronBeamSynthesisModification2022}, while stabilizing the integrity of the underlying oxide structure. In contrast, such B$_2$O$_3$ scales have not been reported in boron-doped alloys, where experimental studies instead observe boron co-segregation with Cr, Fe, and Mo and the formation of boride- or boro-oxide-rich motifs~\cite{peruzzoHightemperatureOxidationSintered2017, leeEffectBoronCorrosion2007a, zhangEvolutionMicrostructureProperties2016a}.

Beyond the direct effects of B--O coordination, the \BO\ system exhibited enrichment of key metallic elements at the oxidized free surface, a result attributed to boron's strong chemical ordering with Cr, Mo, and Fe. This redistribution suggests that boron not only contributes to passivation through its own oxide formation, but also reshapes the local chemical environment in ways that promote the nucleation and stabilization of new oxide structures. Comparative studies on Fe-based amorphous composite coatings~\cite{sharmaHVOFDepositionComparative2023} have shown that boron incorporation enhances fracture toughness, while co-alloying with Mo and Cr improves porosity and corrosion resistance, respectively. Similar findings on Mo--B coatings demonstrated that co-enrichment leads to the formation of Fe- and Mo-based borides, resulting in significantly improved microhardness and superior corrosion resistance under aggressive chlorine exposure~\cite{bartkowskaMicrostructureMicrohardnessCorrosion2020}. It is worth noting that this beneficial behavior stands in contrast to Mo’s standalone oxidation pathway, where MoO$_3$ is volatile and known to degrade oxide scale integrity over time~\cite{qinRoleVolatilizationMolybdenum2017, liEffectMolybdenumCyclic2023}. In the presence of boron, this pathway appears suppressed. Oxidation instead proceeds through the formation of dense, boride-rich surface layers that reduce volatility, enhance surface adherence, and improve long-term corrosion resistance. This behavior is not unique to Mo-based coatings. In boron-doped steels~\cite{peruzzoHightemperatureOxidationSintered2017} and high-entropy alloys~\cite{leeEffectBoronCorrosion2007a, zhangEvolutionMicrostructureProperties2016a}, oxidation often proceeds through the formation of surface-localized borides rather than thick oxide films. These findings suggest that boron actively redirects the oxidation pathway toward stable, low-volatility boride networks, thereby limiting scale growth and improving interfacial stability.

Additionally, diffusion studies involving simultaneous saturation of boron, Cr, and Ti have shown that the resulting surface layers exhibit high hardness, phase selectivity, and compositional depth gradients, depending on the base alloy~\cite{guryevInfluenceChemicalComposition2023}. These findings support the interpretation that boron enables complex interactions with neighboring elements, promoting tunable chemical and mechanical behavior at alloy surfaces. Further evidence comes from boron--Cr coatings, which form layered, boride-rich architectures with enhanced mechanical performance and interfacial stability~\cite{shyrokovSpecificFeaturesFormation2011}. These coatings contained FeB, Fe$_2$B, and mixed Cr--Fe borides that developed into continuous surface layers. Compared to monoboride structures, these phases exhibited reduced brittleness and higher hardness. Notably, the co-diffusion of B and Cr promoted uniform layer formation which is consistent with the early-stage co-enrichment behavior observed at the oxidized free surface in the \BO\ system. Finally, boron's role in establishing oxidation-resistant surface chemistries is evident in Mo--Si--B coatings. In these systems, oxidation led to the formation of a two-layer scale: an upper amorphous aluminoborosilicate layer and a lower crystalline SiO$_2$ layer~\cite{jinMicrostructureHightemperatureOxidation2023}. The boron-rich top layer significantly suppressed oxidation rates by accelerating the formation of protective oxides and enhancing microstructural refinement through improved elemental diffusion. Altogether, these experimental results reinforce the simulation-based finding that boron drives co-enrichment with transition metals at early oxidation stages. They also provide insight into how these initial segregation and coordination motifs may govern the long-term development of chemically complex, mechanically robust surface oxides.

At the \BO\ surface, hMCMD simulations reveal the emergence of complex, polymerized oxide structures with glass-like connectivity, extending far beyond simple adsorption or binary passivation. Boron incorporates into the surface region, forming BO$_3$ trigonal units that link with metal--oxygen chains. For example, boron terminates chains of SiO$_3$ and SiO$_4$ tetrahedra, forming extended B--O--Si linkages. These polymerized motifs resemble structural features of borosilicate glasses, suggesting that boron not only modifies local bonding environments but also facilitates the formation of chemically durable, networked surface structures. In parallel, extended domains of Nb$_2$O$_5$ clusters and Cr--O--Cr chains emerge, forming early-stage oxide clusters with multimetal connectivity. Planar BO$_3$ units appear embedded within or at the interfaces of Nb--O networks, indicating that boron also contributes to mixed-metal oxide polymerization beyond silicate motifs. This behavior highlights boron’s versatility as a network former, capable of integrating into multiple oxide chemistries to enhance connectivity and structural rigidity. The presence of trigonal BO$_3$ units across diverse oxide domains underscores boron’s dual role as both a local modifier and a global network former. Collectively, the coexistence of Nb-, Cr-, B-, and Si-based oxides suggests that the boron-doped surface is primed to evolve into a chemically heterogeneous, structurally robust oxide network. While these simulations capture only the early stages of surface evolution, the observed motif formation provides a mechanistic basis for boron’s ability to promote self-passivating, corrosion-resistant surface architectures. These observations are consistent with experimentally observed improvements in the oxidation resistance of boron-doped Ni-based superalloys.

\subsubsection*{Dynamic Segregation and Grain Boundary Composition}
\begin{figure}[H]
    \centering
    \includegraphics[width=\linewidth]{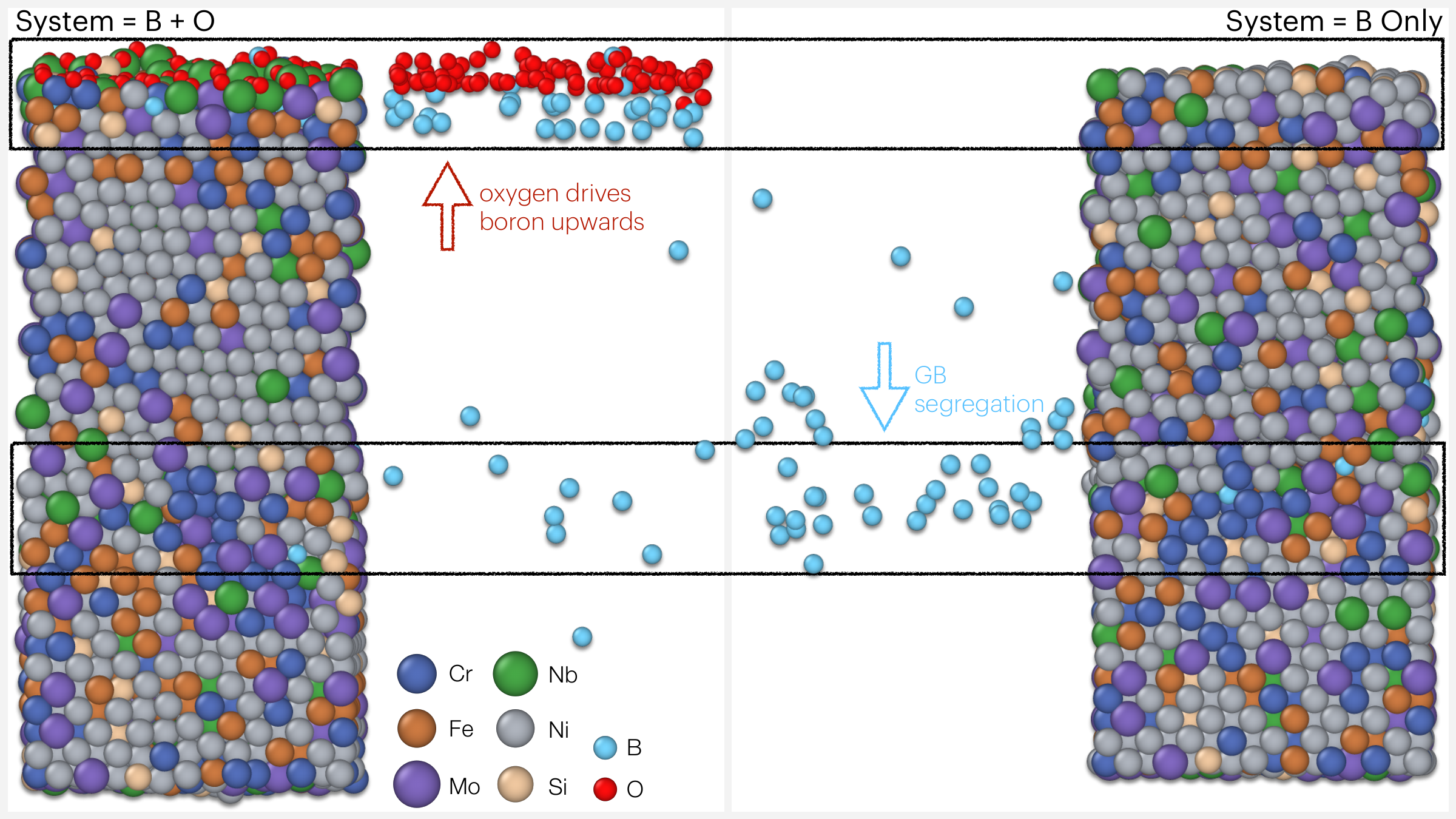}
    \caption{Side-by-side comparison of boron distribution in the mixed-metal lattice with and without oxygen. In the presence of O (left), boron strongly accumulates at the free surface, while in the absence of O (right), boron preferentially segregates to the grain boundary.
}
    \label{fig:segregation}
\end{figure}

Following Fig.~\ref{fig:ordering}a and Fig.~\ref{fig:segregation}, the simulations reveal that boron exhibits a strong tendency to segregate to the free surface in the presence of oxygen, where it forms mixed boro-oxides and clusters within the subsurface layers. In contrast, relatively limited segregation to the grain boundary is observed in the \BO\ system compared to the \Bonly\ case. Interestingly, boron does not predominantly form crystalline oxide structures at the free surface. Instead, it appears thermodynamically stabilized just beneath the surface, where it co-segregates with transition metals. This behavior is supported by experimental observations. Peruzzo et al.~\cite{peruzzoHightemperatureOxidationSintered2017} reported that boron-doped stainless steels exhibited significantly reduced oxidation rates and minimal oxide scale detachment, attributed to densification from liquid-phase sintering and complex surface interactions among Cr, Fe, and boron. EDS analysis revealed pronounced co-segregation of boron with Cr in intergranular regions at the surface. One of the earliest investigations into boron-doped high-entropy alloys~\cite{leeEffectBoronCorrosion2007a} reported the formation of (Cr,Fe,Co)-borides, accompanied by Cr depletion outside these boride phases and improved surface corrosion resistance. More recently, Zhang et al.~\cite{zhangEvolutionMicrostructureProperties2016a} confirmed similar behavior, identifying (Cr,Fe)$_2$-borides at the oxidized surface that were directly correlated with enhanced corrosion resistance.

The close agreement between these experimental findings and the simulated surface segregation pathways lends strong support to the predictive capability of the MC routine. In particular, the emergence of B--Cr association motifs under oxidizing conditions, captured without any bias toward interface placement, demonstrates that atomistic simulations can resolve subtle, thermodynamically driven segregation phenomena that govern long-term surface evolution. These results underscore the importance of accounting for dopant redistribution not only at grain boundaries but also at free surfaces, where oxidation resistance is often initiated or compromised. They further highlight the chemical cross-talk between light interstitials and metallic species during oxidation, wherein local coordination environments and segregation tendencies are jointly modulated by the presence of reactive elements such as oxygen, carbon, or boron.

Focusing on GB chemical composition, the data reveal clear trends in how different elements respond to oxygen and boron, as shown in Fig.~\ref{fig:ordering}a. In the \clean\ system, the GB exhibits modest enrichment in Cr +0.58\%, Mo +1.46\%, Nb +1.56\%, and Si +1.92\%, while Fe and Ni are depleted. This passive redistribution reflects the absence of strong driving forces in a chemically inert environment. Exposure to oxygen alone in the \Oonly\ system dramatically alters this profile. Chromium undergoes substantial depletion --2.56\%, while Mo and Si show increased enrichment of +4.42\% and +2.86\%, respectively. Nb enrichment is reduced to +0.70\%, compared to +1.56\% in the \clean\ system. These shifts arise from the surface segregation of Cr and Nb in response to oxygen adsorption, which depletes their availability at the GB and allows Mo and Si to further enrich. This highlights the dynamic chemical partitioning that emerges between competing interfacial environments, particularly when one is decorated with light interstitials.

In the \Bonly\ system, boron segregates strongly to the GB at +3.80\%, accompanied by pronounced enrichment of Cr +5.00\% and Mo +3.40\%. Nb enrichment remains comparable at +1.40\%, while Fe depletion is nearly neutral at --0.08\%, and Si shows only slight enrichment at +0.80\%, suggesting limited B--Si interaction. This condition also exhibits the most severe Ni depletion across all systems, with a reduction of --10.52\%, indicating that boron drives substantial internal redistribution and displaces Ni from interfacial sites. These compositional shifts are corroborated by the SRO data in Fig.~\ref{fig:ordering}c (right), where negative $\alpha_{\rm BM}$ values indicate strong local clustering between boron and Cr, Fe, Mo, and Nb. Together, the segregation and SRO results underscore boron’s role in reshaping the interfacial chemical environment through the formation of preferentially coordinated, metal-rich motifs. The GB affinity of boron and the resulting compositional trends (e.g., strong ordering between boron with Cr, Mo, and Nb) align well with experimental observations in boron-doped Inconel alloys~\cite{tekogluStrengtheningAdditivelyManufactured2023, tekogluMetalMatrixComposite2024b, tekogluSuperiorHightemperatureMechanical2024c, cadelAtomProbeTomography2002, tytkoMicrostructuralEvolutionNibased2012}. Moreover, the \Bonly\ system exhibits a total energy 275~eV lower than that of the \clean\ system, indicating substantial interfacial stabilization. This energetic reduction reflects boron’s favorable incorporation at the GB and its capacity to lower overall system energy. Previous work has similarly shown that boron segregation to Cr--Ni grain boundaries continuously reduces both the grain boundary energy and segregation energy with increasing boron content~\cite{DOLEZAL2025121221}, further affirming boron’s stabilizing role.

When both boron and oxygen are present in the \BO\ system, GB boron levels drop markedly to +0.20\% due to strong surface segregation. Residual boron modestly mitigates Cr depletion compared to the \Oonly\ case, shifting from --2.56\% to --0.12\%, while Mo and Si remain moderately enriched at +3.25\% and +1.54\%, respectively. This reflects the continued chemical competition between surface and GB sites and suggests that even small amounts of interfacial boron can influence elemental partitioning under oxidizing conditions. Overall, these trends reinforce boron’s role as a potent interfacial modifier capable of stabilizing grain boundaries, redirecting elemental partitioning, and competing effectively with oxygen for segregation sites.

\subsubsection*{Short-Range Ordering Away from and at the Free Surface}

Short-range ordering away from the free surface, shown in Fig.~\ref{fig:ordering}c (right), reveals that the M--M$'$ SRO values across all systems (circle, square, and triangle markers) remain remarkably similar to those of the \clean\ reference (x markers). This consistency indicates that the core metallic bonding network is largely preserved in the interior of the system. It reinforces the interpretation that structural reconfigurations in the oxidized and doped cases are driven primarily by surface-localized chemistry and do not reflect global disruption of the internal metallic lattice. These oxidation-induced perturbations appear to penetrate no more than four atomic layers from the surface, beyond which the SRO remains effectively unchanged.

In contrast, pronounced deviations in M--M$'$ SRO values are observed at the free surface, reflecting the volatility of local chemical environments in response to reactive interstitials. Oxygen exposure alone (square markers) enhances ordering in pairs such as Cr--Nb and Fe--Mo, while the addition of boron (circle markers) introduces further disruptions. In particular, Cr--Cr ordering is suppressed in the \BO\ system, likely due to the formation of Cr--B--Cr bridging motifs. Similar bridging interactions are evident in Fe--Mo and Mo--Mo pairs, where the presence of subsurface boron facilitates mixed-element linkages such as Fe--B--Mo and Mo--B--Mo. These B-centered bridges are enabled by subsurface enrichment of Cr, Fe, and Mo just below the oxidized, Nb-rich surface layer, providing the atomic proximity required for multi-element coordination. Both the \Oonly\ and \BO\ systems show evidence of Nb--Nb clustering at the free surface, a feature absent in the \clean\ reference. Together, these trends underscore the surface as a chemically dynamic region, where short-range ordering responds rapidly to the presence of reactive light interstitials. This behavior stands in sharp contrast to the relative structural stability of the interior lattice, which remains largely unaffected by surface oxidation or dopant incorporation.

\subsection*{Chlorine Disruption of Oxide Networks}

\begin{figure}[H]
    \centering
    \includegraphics[width=\linewidth]{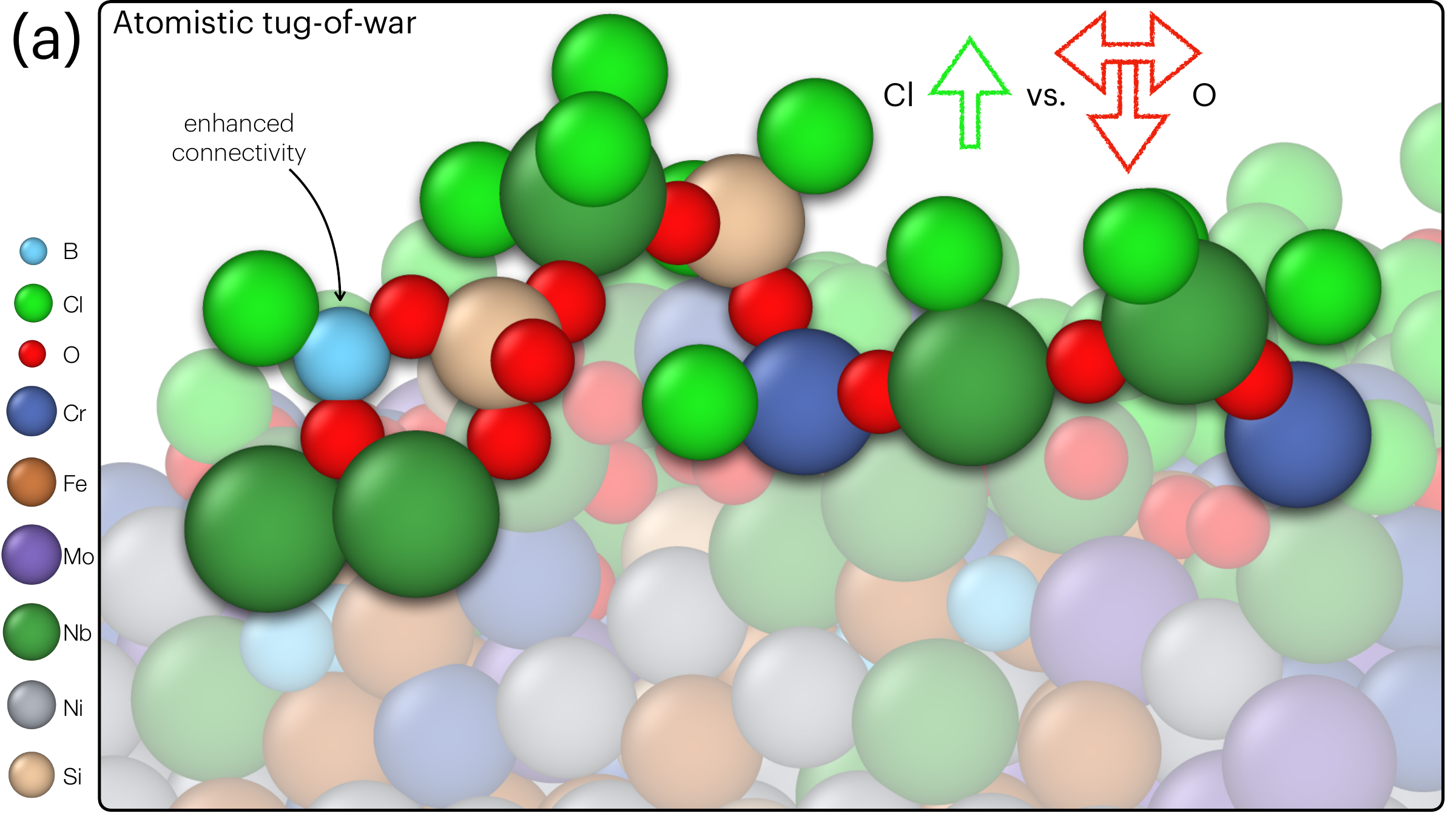}
    \begin{minipage}[t]{0.325\linewidth}
        \centering
        \includegraphics[width=\linewidth]{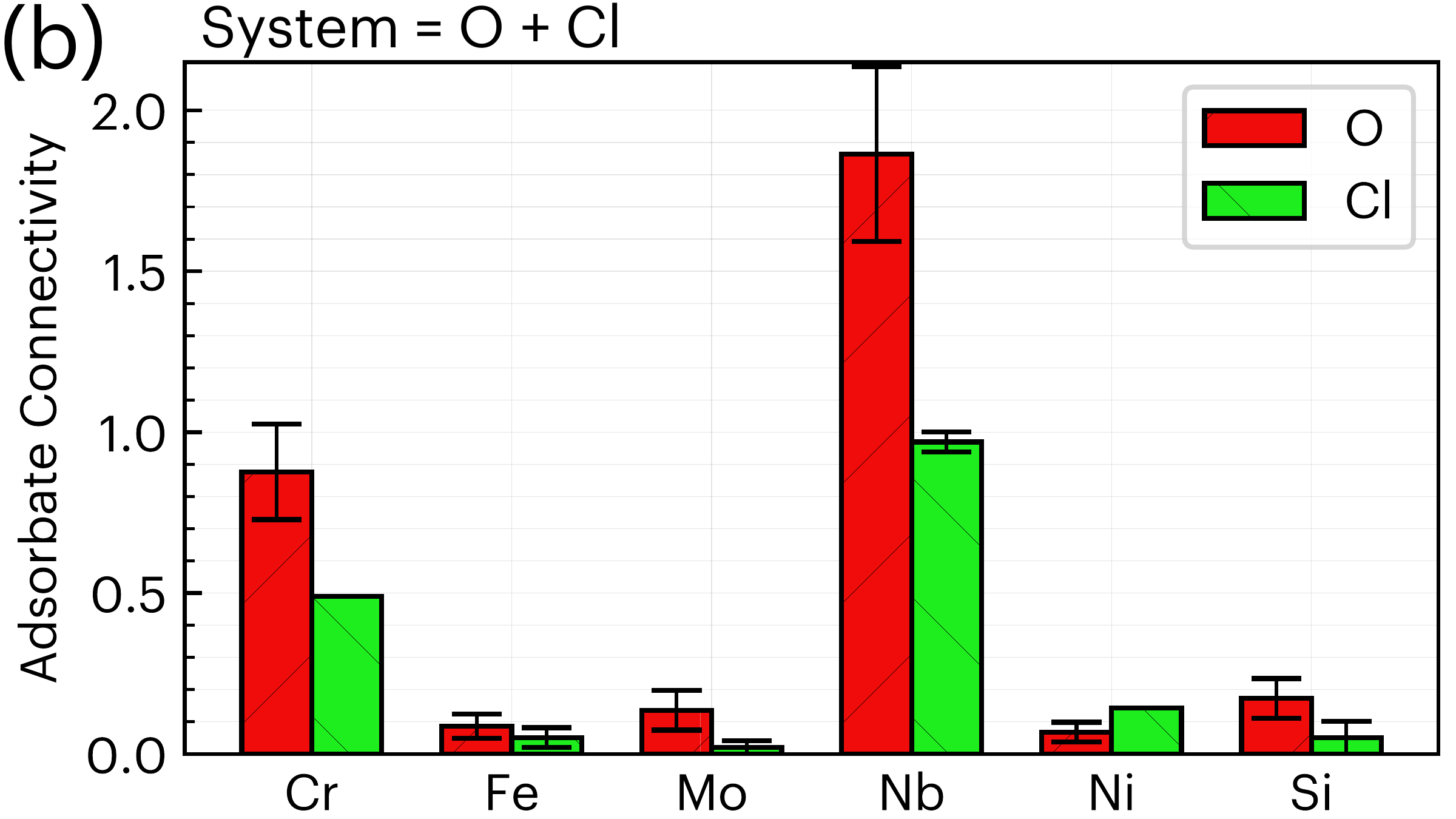}
    \end{minipage}
    \hfill
    \begin{minipage}[t]{0.325\linewidth}
        \centering
        \includegraphics[width=\linewidth]{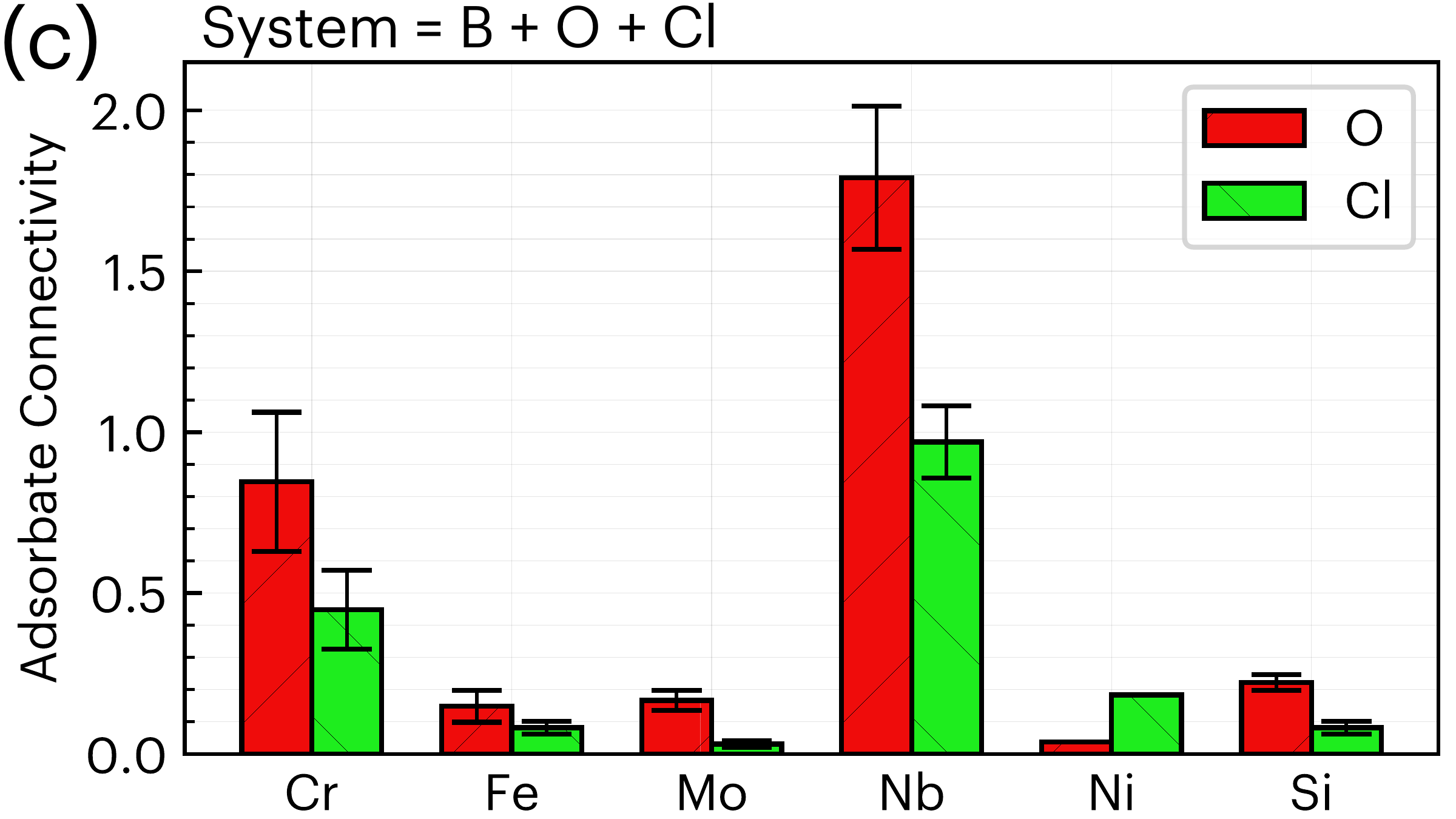}
    \end{minipage}
    \begin{minipage}[t]{0.325\linewidth}
        \centering
        \includegraphics[width=\linewidth]{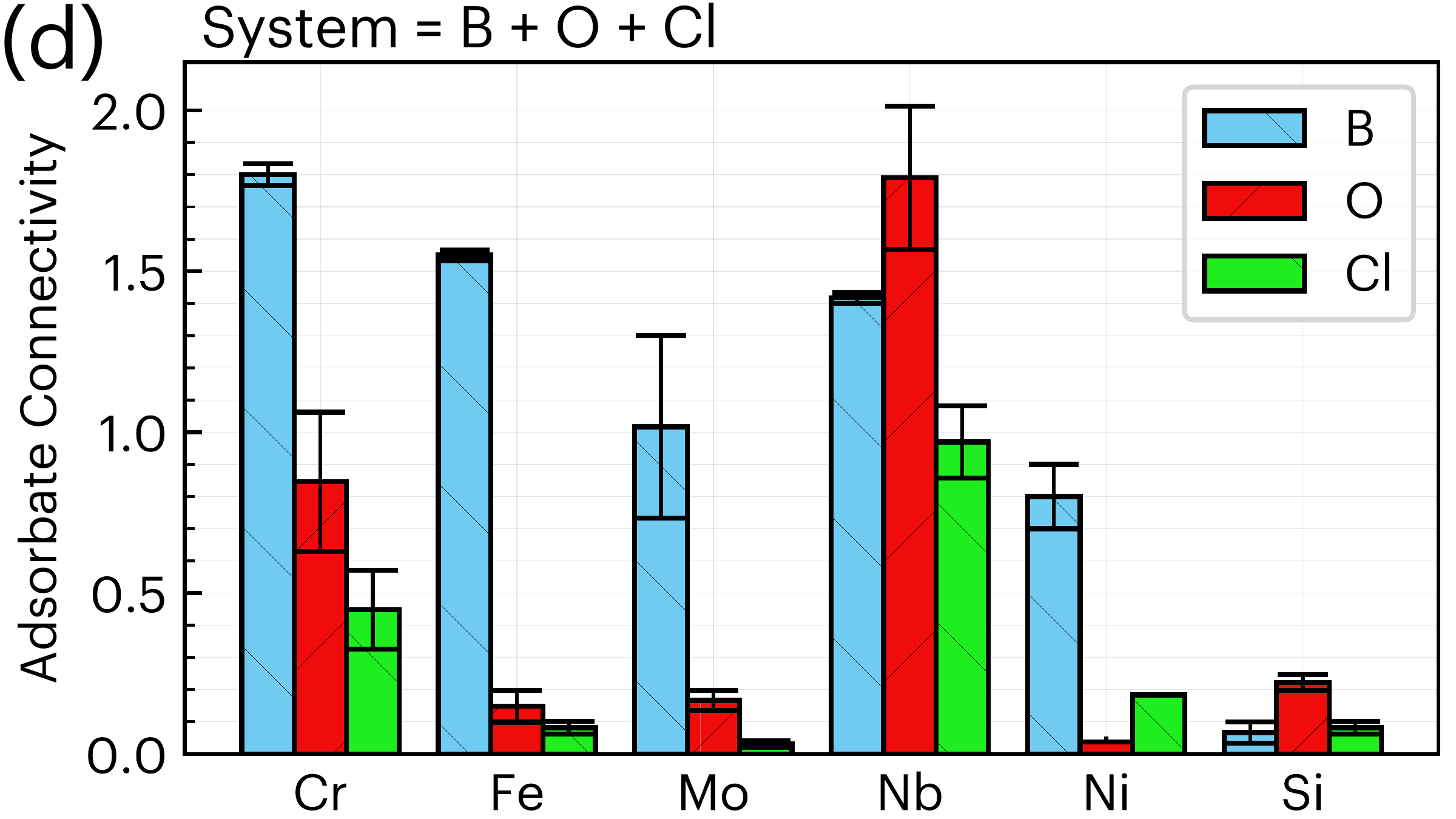}
    \end{minipage}
    \caption{(a) Final MD snapshot of the \BO\ + Cl system illustrates the atomistic tug-of-war between adsorbed chlorine atoms and the underlying metal-oxide network. Chlorine pulls upward, while the oxygen-stabilized oxide anchors the structure through lateral and downward coordination. Animations show that B-O clusters reduce atomic mobility in nearby metal atoms, resulting in a more rigid oxide. (b-d) Average number of metal neighbors per oxygen, chlorine, or boron atom, resolved by metal type, for (b) the O + Cl system, (c) the \BO\ + Cl system, and (d) including the boron atoms in the \BO\ + Cl system. Bars represent the mean M--O, M--Cl, or M--B coordination per light atom, normalized by the number of oxygen, chlorine, or boron atoms, respectively. Error bars reflect standard deviations across two independent simulations.}
    \label{fig:tugowar}
\end{figure}

\subsubsection*{Changes in Surface Composition}

Changes in surface chemistry following chlorine adsorption, MC equilibration, and a 20~ps $NVT$ MD simulation are shown in Fig.~\ref{fig:ordering}a. Both the \Oonly\ and \BO\ systems exhibit further Nb enrichment after chlorine exposure, with a more pronounced increase in the \Oonly\ surface compared to a more modest rise in \BO. While Cr content increases at the O + Cl surface, it decreases slightly in the \BO\ + Cl system, accompanied by corresponding increases in Fe and Mo concentrations. This suggests that chlorine preferentially coordinates with Fe and Mo in the presence of boron, altering the surface segregation hierarchy.

In both systems, Ni becomes more strongly depleted after chlorine exposure. A subtle increase in surface boron content is also observed in the \BO\ system, indicating some degree of retention despite competition with chlorine. These chemical alterations correspond with changes in interstitial--metal SRO (i--M SRO), shown in Fig.~\ref{fig:ordering}c (left). For instance, $\alpha_{\rm ONb}$ becomes less negative in both systems, coinciding with strong Cl--Nb association as captured by more negative $\alpha_{\rm ClNb}$ values. A similar effect is observed for $\alpha_{\rm OCr}$, where chlorine coordination with Cr disrupts the preexisting O--Cr ordering. These trends reflect chlorine’s capacity to destabilize oxygen-driven coordination environments, particularly among metals with high oxygen affinity.

\subsubsection*{Dynamic Response of the Oxide Network}

To probe the surface network's response to chlorine attack at elevated temperature, Fig.~\ref{fig:tugowar}a presents the final configuration of the MD simulation at 1073~K for the \BO\ + Cl system. This snapshot illustrates the atomistic ``tug-of-war'' between adsorbed chlorine atoms and the underlying mixed-metal oxide network. Animation of the trajectory reveals that M--O--M linkages remain exceptionally rigid, maintaining stable bond angles and acting as structural anchors for exposed surface atoms. Oxygen atoms exert a stabilizing lateral and downward pull, reinforcing the connectivity of the oxide network and resisting deformation. In contrast, chlorine atoms adsorbed atop metal centers undergo frequent bond angle fluctuations and impose an upward pull, disrupting local coordination as they respond to both the rigidity of the oxide framework and interactions with neighboring Cl species. Also evident is a BO$_3$ trigonal unit in which a chlorine atom has substituted for oxygen, highlighting a localized disruption within an otherwise stable boro-oxide domain. These BO$_3$ groups are distributed across the surface and mark regions of exceptional rigidity, with minimal displacement in their surrounding oxide environment (RMSD = 0.94~\AA). Strongly coordinated Nb centers, particularly those in NbO$_4$ and NbO$_5$ configurations, remain firmly embedded within the network, resisting displacement and preserving local bonding topologies throughout the simulation.

A key stabilization mechanism emerges uniquely in the \BO\ + Cl system: an anchoring effect exerted by subsurface boron atoms. These boron species form strong vertical bonds with overlying metal atoms, acting as buried anchors that suppress atomic motion at the surface. Importantly, Cr atoms directly above subsurface boron sites exhibit dramatically reduced oscillations throughout the simulation, maintaining stable coordination geometries even under aggressive chlorine exposure. This anchoring behavior is evident in Fig.~\ref{fig:tugowar}d, which highlights strong B--Cr coordination at the surface (blue bars). In contrast, Cr atoms lacking subsurface boron support, and coordinated only to surface oxygen or chlorine, exhibit markedly higher mobility. These unanchored Cr atoms form Cr--Cl adatoms, with chlorine adsorbed atop Cr, that skim across the surface while forming transient Cr--O bonds that fail to stabilize them. The resulting Cr--Cl species oscillate between brief coordination and detachment, reflecting a breakdown in local oxide connectivity and mechanical cohesion. This boron-mediated anchoring effect emerged consistently across simulations and aligns with experimental observations of boron’s role in enhancing surface corrosion resistance~\cite{leeEffectBoronCorrosion2007a, zhangEvolutionMicrostructureProperties2016a}. Both studies report the formation of Cr and Fe-based borides at alloy surfaces, an observation that is corroborated by the strong B--Cr and B--Fe bonding revealed in the surface connectivity analysis (Fig.~\ref{fig:tugowar}d). These results underscore boron’s dual role: reinforcing the oxide network through BO$_3$ motifs and anchoring otherwise labile metal species, particularly Cr, to suppress chlorine-driven degradation.

Taken together, this dynamic interplay offers atomistic insight into how a boron- and Nb-enriched oxide network can ``weather the storm'' of persistent halogen attack. Importantly, this surface anchoring mechanism has not been previously reported in atomistic simulations. It reveals an unrecognized role for light interstitials, not only as network formers or modifiers, but also as surface stabilizers that actively suppress metal volatility under corrosive conditions. This finding suggests that the deliberate placement of boron near surface regions could serve as a design strategy for enhancing halogen resistance in advanced alloys, offering a new perspective on how interstitial chemistry governs degradation resistance at the atomic scale.

\subsubsection*{Network Connectivity and Chemical Durability}

A more detailed view of preferred coordination behavior is presented in Fig.~\ref{fig:tugowar}b-d, which report the average number of metal neighbors coordinated to each oxygen, chlorine, or boron atom, normalized by the total number of oxygen, chlorine, or boron atoms, respectively. This metric serves as a measure of the ``connectivity'' between surface adsorbates and the underlying metal network. In Fig.~\ref{fig:tugowar}b and c, oxygen and chlorine connectivity trends are shown. Oxygen tends to form extended M--O--M chains and oxide clusters, whereas chlorine exhibits more site-specific coordination, often binding atop individual metal centers. Such discrete coordination is commonly associated with early-stage compound formation or incipient desorption. In this context, the red bars in Fig.~\ref{fig:tugowar}b-d reflect the connectivity of the oxide network: O--Nb linkages dominate across both the \Oonly\ and \BO\ surfaces, followed by O--Cr. There is a noticeable increase in oxygen connectivity with Cr, Fe, Mo, and Nb after chlorine exposure, indicating a reinforcement of M--O--M chains in response to corrosive attack. This suggests that the oxide network adapts dynamically to maintain structural cohesion. In the \BO\ system, oxygen connectivity with Fe and Mo is particularly enhanced, reflecting their boron-driven enrichment in the subsurface region and subsequent migration to active coordination sites at the surface. In contrast, chlorine connectivity highlights the metals most affected by halogen attack. Nb and Cr show the strongest chlorine coordination, followed by Ni and Fe. These results emphasize the duality of surface response: oxygen facilitates network formation and rigidity, while chlorine targets specific metal centers, potentially undermining local connectivity and stability. 

Fig.~\ref{fig:tugowar}d presents the addition of boron connectivity in the \BO\ + Cl system. It reveals that boron is most strongly bonded to Cr and Fe at the surface, consistent with experimental observations of Cr- and Fe-based boride formation \cite{peruzzoHightemperatureOxidationSintered2017, leeEffectBoronCorrosion2007a,zhangEvolutionMicrostructureProperties2016a}. This surface-specific coordination behavior supports the interpretation that boron plays an anchoring and stabilizing role under corrosive conditions, complementing the trends observed for oxygen and chlorine. For example, the strong preference for Fe coordination appears to be unique to the surface environment. In contrast, boron within the grain boundary of the \Bonly\ system is more strongly ordered with Cr, Mo, and Nb, as shown in Fig.~\ref{fig:ordering}c. This distinction highlights the influence of local chemical and geometric constraints and suggests that competitive interstitial environments, such as surfaces with B, O, and Cl, may drive new modes of chemical ordering not observed in isolated bulk or GB systems.

Further surface analysis revealed that chlorine coordination is highly dependent on the local oxygen environment and the vertical displacement of metal atoms relative to the oxide network. Specifically, metal atoms displaced upward and out of the surface plane become increasingly susceptible to chlorine adsorption. In contrast, metal atoms that remain coplanar with the oxide network and are fully coordinated with bridging oxygen atoms are effectively protected from chlorine attack. For example, Nb atoms embedded within the planar oxide structure with NbO$_4$ coordination exhibited no chlorine adsorption. In contrast, out-of-plane Nb atoms with lower oxygen coordination (e.g., NbO$_2$) supported multiple chlorine attachments, often forming NbCl$_3$ units. Bridge-site geometries such as Ni--Cl--Nb, Nb--Cl--Nb, and Cr--Cl--Cr were observed in surface-planar regions where metal atoms exhibited partial oxygen coordination. Vertical displacements of 1--2~\AA{} from the oxide surface correlated with increased chlorine affinity, frequently resulting in atop coordination with two or more chlorine atoms. More extreme displacements (3--6~\AA{}) further enhanced chlorine binding, with larger chlorine clusters forming around under-coordinated sites due to increased accessible surface area. This phenomenon is illustrated in the top panel of Fig.~\ref{fig:tugowar}, where an Nb--O--Si segment of the metal-oxide network is visibly elevated out of plane. The reduced oxygen coordination of these atoms enabled the formation of both NbCl$_3$ and SiCl$_2$ species at the exposed interface. 

A consistent trend emerged: chlorine attraction increased with vertical displacement and decreased oxygen saturation. Out-of-plane NbO$_2$ structures consistently supported Cl$_3$ binding, whereas NbO$_3$ and NbO$_4$ sites coordinated only one or zero chlorine atoms, respectively. Similar patterns were observed for silicon: out-of-plane SiO$_2$ hosted Cl$_2$, SiO$_3$ supported a single chlorine atom, and fully coordinated SiO$_4$ units resisted chlorine entirely, even when displaced out of plane. For Cr, out-of-plane CrO$_3$ bound a single Cl atom, while CrO and CrO$_2$ supported Cl$_2$ coordination. In contrast, Mo, Ni, and Fe atoms were not observed in elevated positions and supported only bridging-type chlorine configurations (e.g., M--Cl--M). These results demonstrate that chlorine coordination is governed by a combination of vertical displacement and oxygen saturation, reinforcing the importance of oxide network rigidity in resisting halogen-induced attack.

To further quantify how structural rigidity varies across metal centers, the distribution of M--O--M bond angles was analyzed to assess differences in bridging geometry and network coherence. Nb--O--Nb angles were consistently wide (110.2$^\circ$--112.1$^\circ$), indicative of an open bridging geometry conducive to extended oxide network formation. In contrast, Cr--O--Cr angles were narrower (95.1$^\circ$--98.2$^\circ$), reflecting a more compact or distorted environment, likely associated with localized rather than extended bridging. In comparison, Cl--M--X angles (where X = Cl or O) represent terminal, non-bridging coordination. For Cr, these angles ranged from 102.1$^\circ$ to 103.2$^\circ$, while for Nb they were slightly smaller (96.7$^\circ$--100.9$^\circ$), suggesting tighter packing or increased distortion around the larger Nb cation. While M--O--M angles remained relatively stable throughout the MD simulation, the Cl--M--X angles fluctuated continuously, reflecting the dynamic, loosely coordinated nature of chlorine interactions with the surface and their sensitivity to local atomic rearrangements. Collectively, these angular distributions highlight a clear structural distinction: oxygen enables rigid, wide-angle bridging (M--O--M) that supports network connectivity, whereas chlorine favors dynamic, terminal coordination (Cl--M--X) that disrupts coherence and resists polymerization.

The coordination trends underscore the importance of connectivity not only in resisting chlorine-induced disruption, but also in maintaining oxide scale cohesion. In particular, high oxygen coordination around Nb atoms, such as in NbO$_4$ and NbO$_5$ configurations, plays a critical role in fortifying the oxide network. These configurations are chemically stable, resisting both chlorine adsorption and surface displacement, and are structurally significant as anchors within the evolving oxide layer. Experimental studies have shown that fully coordinated Nb$_2$O$_5$ sublayers enhance the adhesion of chromia-based surface scales to the underlying alloy, forming dense, coherent interfacial layers that reduce the risk of spallation and delamination \cite{yeInfluenceNbAddition2021, haInvestigatingOxideFormation2024, gengOxidationBehaviorAlloy2007}. In the present simulations, highly coordinated Nb centers serve as rigidity nodes, contributing simultaneously to oxide network integrity and long-term surface passivation. This connection between local oxygen coordination and macroscopic interfacial stability reveals a broader design principle: the durability of protective oxide scales depends not only on chemical composition, but also on the spatial and topological continuity of the underlying bonding network.

Additionally, these observations provide a mechanistic link to early-stage basic fluxing behavior~\cite{youngChapter8Corrosion2016, youngChapter12Corrosion2016}. Under-coordinated Nb atoms, particularly those displaced from the oxide network and exhibiting NbO$_2$-type configurations, act as electronically unsaturated Lewis acid sites. Their reduced oxygen coordination decreases crystal field stabilization and exposes open valences, enhancing chemical reactivity. These sites preferentially accommodate multiple Cl adsorbates, as seen in the formation of NbCl$_3$, and serve as nucleation points for volatile halide production. In doing so, they disrupt local oxide connectivity and diminish the structural coherence of the passivating layer. A complementary degradation pathway emerges in the behavior of Cr--Cl adatoms, which exhibit high lateral mobility and fail to incorporate into the oxide network. These species oscillate between transient Cr--O bonding and detachment, suggesting a frustrated chemisorption state that agitates the surrounding structure without contributing to passivation. Over time, both degradation mechanisms, Nb-centered halide formation and Cr-induced dynamic disruption, undermine the oxide matrix and facilitate further chlorine ingress. Thermodynamically, these processes promote the formation of volatile chlorides (e.g., NbCl$_5$, CrCl$_3$), especially in regions where oxygen coordination is compromised. The atomistically resolved interplay between oxygen coordination, vertical displacement, and halogen mobility thus mirrors experimental observations of basic fluxing, wherein chemically and structurally weakened regions are preferentially attacked. These findings reinforce the critical role of a highly coordinated, polymerized oxide interface capable of distributing stress and resisting local adsorption as a prerequisite for maintaining surface integrity in halogen-rich environments.

Taken together, these atomistic insights into interfacial evolution, network resilience, and halogen-driven disruption align closely with experimental observations of enhanced corrosion resistance in complex alloy systems. Two boron-driven stabilization mechanisms emerged from the simulations. First, rigid, polymerized surface oxide motifs, particularly those anchored by BO$_3$ and Nb--O units, enabled lateral strain distribution and suppressed chlorine-induced structural breakdown. These findings are consistent with experimental reports showing that boron migrates to the surface during oxidation and contributes to the formation of protective, B-rich oxide caps~\cite{glechnerInfluenceNonmetalSpecies2021, yunactiReplacingToxicHard2023}, while also enhancing resistance to Cl-induced pitting~\cite{yushkovElectronBeamSynthesisModification2022}. Second, a newly identified subsurface anchoring mechanism revealed that boron atoms embedded just beneath the surface formed vertical linkages with overlying Cr atoms, significantly restricting their mobility and suppressing volatilization under chlorine exposure. This anchoring behavior is strongly supported by experimental studies of boron-doped steels and high-entropy alloys~\cite{peruzzoHightemperatureOxidationSintered2017, leeEffectBoronCorrosion2007a, zhangEvolutionMicrostructureProperties2016a}, which consistently report Cr co-segregation with boron and the formation of boride-stabilized surface phases that enhance corrosion resistance. 

Beyond boron, Nb also played a critical stabilizing role. Experimental studies have shown that Nb promotes Nb$_2$O$_5$ formation, reduces film defects, and enhances Cr$_2$O$_3$ scale adherence under halogen-rich environments~\cite{xiaoEffectNiNb2025, wengInfluenceNbHot2014}. These effects were mirrored in the present simulations, where Nb-rich regions maintained structural integrity during chlorine exposure and resisted fluxing. The observed Cl--Nb affinity is further supported by first-principles calculations, which show strong Nb--Cl binding and resistance to desorption up to NbCl$_5$ in Ni--Nb--W environments~\cite{dolezalFirstprinciplesStudyEarlystage2023}. Altogether, the simulations provide mechanistic validation for the experimentally reported benefits of light interstitial doping and inclusion of Nb. They underscore the importance of coordinated oxide network formation, controlled elemental redistribution, and engineered interfacial architectures in designing halogen-resistant surface chemistries. Notably, boronized alloys containing Cr and Mo have demonstrated improved long-term corrosion resistance and stability~\cite{bartkowskaMicrostructureMicrohardnessCorrosion2020, peruzzoHightemperatureOxidationSintered2017, leeEffectBoronCorrosion2007a, zhangEvolutionMicrostructureProperties2016a, shyrokovSpecificFeaturesFormation2011}, echoing the polymerized M--O--M frameworks and surface anchoring mechanisms captured in these simulations.

\subsection*{Thermodynamic Stability of Surface Oxidation and Degradation}

To confirm the thermodynamic viability of the oxidized and chlorine-exposed surface configurations examined in this study, the Gibbs free energy change ($\Delta G$) associated with oxygen and chlorine adsorption at 1073~K was evaluated. As shown in Fig.~S5 of the Supplemental Materials, both adsorption processes were thermodynamically favorable on the \Oonly\ and \BO\ surfaces, exhibiting negative $\Delta G$ values across the full range of chemical potentials considered. These trends affirm that the observed chemical rearrangements and structural motif formations arise under thermodynamically downhill conditions. Accordingly, the surface states analyzed through MD simulations and SRO metrics represent physically realizable configurations within realistic operating environments.

\section{Conclusion}

This study demonstrates how light interstitials, boron and oxygen, alter chemical distribution at reactive interfaces and influence one another to drive new interfacial chemistries. Hybrid MC/MD simulations revealed that boron reinforces surface stability not only through BO$_3$ trigonal motifs embedded within mixed-metal oxide networks, but also by driving co-enrichment with Cr, Fe, and Mo in the subsurface region, enhancing M--B--M connectivity. This led primarily to boron-anchored Cr, which exhibited enhanced resistance against chlorine bonding and subsequent desorption. The resulting anchoring mechanism suppressed atomic mobility and mitigated chlorine-induced degradation, representing a previously unreported phenomenon revealed by these simulations. Crucially, boron exhibited environment-sensitive segregation: it preferred the grain boundary and enriched it with Cr, Mo, and Nb in the absence of oxygen, but strongly segregated to the subsurface region of the free surface upon oxidation. Oxygen adsorption produced a parallel effect by driving the formation of new surface chemistry with high enrichment of Cr and Nb. Together, these results highlight how cross-talk between light interstitials reshapes alloy-element distribution and drives the emergence of chemically resilient oxide architectures.

Niobium emerged as a key stabilizing agent under chlorine-rich conditions. Niobium maintained robust Nb--O coordination and supported wide Nb--O--Nb bond angles, providing structural continuity while accommodating chlorine adsorption without compromising network integrity. These Nb-rich regions served as chemical and structural buffers within the evolving oxide framework. High-temperature MD simulations captured the dynamic competition between chlorine-induced disruption and oxygen-mediated stabilization. While oxygen-based Cr--O bonds alone were insufficient to prevent Cr--Cl adatom formation, Cr atoms reinforced by subsurface Cr--B linkages exhibited markedly reduced mobility and resisted chlorine pull-off. This boron-mediated stabilization of otherwise labile Cr centers aligns with experimental reports of suppressed chromium volatilization in boron-doped alloys. Regions enriched in Nb and BO$_3$ motifs further anchored the surface and preserved oxide network integrity, while unanchored Cr sites remained susceptible to Cl-induced disruption.

By resolving spatially distinct segregation behaviors, coordination motifs, and dynamic degradation pathways, this work establishes a mechanistic foundation for interfacial stabilization in halogenated environments. The identification of surface anchoring, M--B--M co-enrichment, and Nb-centered resilience advances understanding of early-stage degradation and provides actionable insight into how light interstitials mediate alloy-element segregation and drive dynamic early-stage interfacial chemistries. In this way, our atomistic results complement experimental observations of boron co-segregation with Cr, Fe, and Mo, pointing toward design criteria for oxidation- and chlorine-resistant Ni-based superalloys.

\section*{Author Contributions}
\textbf{T.D.D} Writing - original draft, writing - review \& editing, visualization, validation, software, methodology, investigation, formal analysis, data curation. \textbf{R.F.} Project administration and supervision, writing - review \& editing. \textbf{J. Li} Project administration and supervision, writing - review \& editing, funding acquisition. All authors contributed to the conceptualization of this project.

\section*{Data Availability}
The hybrid Monte Carlo Molecular Dynamics (hMCMD) routine is available at \url{https://github.com/tylerdolezal/hybrid_MCMD}. All data generated from this work and post-processing scripts will be provided at \url{https://github.com/tylerdolezal}.

\section*{Acknowledgments}
J. Li acknowledges support from National Science Foundation, USA CMMI-1922206 and DMR-1923976.

\bibliographystyle{ieeetr}

\end{document}